\documentclass[%
 reprint,
nofootinbib,
 amsmath,amssymb,
 aps,
]{revtex4-2}

\usepackage[dvipsnames]{xcolor}

\definecolor{BerlinU1}{HTML}{62AD2D}
\definecolor{BerlinU2}{HTML}{E94D10}
\definecolor{BerlinU3}{HTML}{00A192}
\definecolor{BerlinU9}{HTML}{F18800}

\usepackage{graphicx}
\usepackage{dcolumn}
\usepackage{bm}
\usepackage[normalem]{ulem}
\usepackage{tikz}
\usetikzlibrary{tikzmark}
\usepackage{chemformula}

\usepackage[colorlinks=true]{hyperref}
\hypersetup{
    colorlinks = true,
    linkcolor = BerlinU9,
    urlcolor = BerlinU9,
    linkbordercolor = {white},
    citecolor = BerlinU9,
    }

\usepackage{orcidlink}
\usepackage{pifont}
\newcommand{\cmark}{\textcolor{BerlinU1}{\ding{51}}}%

\newcommand{\BH}{\bullet}
\newcommand{\NS}{\circ}

\begin{document}

\title{Challenging a binary neutron star merger interpretation of GW230529}

\author{Ivan \surname{Markin}\,\orcidlink{0000-0001-5731-1633}$^{1}$}
\email{ivan.markin@uni-potsdam.de}
\author{Anna \surname{Puecher}\,\orcidlink{0000-0003-1357-4348}$^{1}$ } 
\author{Mattia \surname{Bulla}\,\orcidlink{0000-0002-8255-5127}$^{2,3,4}$ }%
\author{Tim \surname{Dietrich}\,\orcidlink{0000-0003-2374-307X}$^{1,5}$}

\affiliation{${}^1$ Institut f\"ur Physik und Astronomie, Universit\"at Potsdam, Haus 28, Karl-Liebknecht-Str. 24/25, 14476, Potsdam, Germany}
\affiliation{${}^2$ Department of Physics and Earth Science, University of Ferrara, via Saragat 1, I-44122 Ferrara, Italy}
\affiliation{${}^3$ INFN, Sezione di Ferrara, via Saragat 1, I-44122 Ferrara, Italy}
\affiliation{${}^4$ INAF, Osservatorio Astronomico d’Abruzzo, via Mentore Maggini snc, 64100 Teramo, Italy}
\affiliation{${}^5$ Max Planck Institute for Gravitational Physics (Albert Einstein Institute), Am M\"uhlenberg 1, Potsdam 14476, Germany}

\date{\today}

\begin{abstract}
GW230529\_181500 represented the first gravitational-wave detection with one of the component objects' mass inferred to lie in the previously hypothesized mass gap between the heaviest neutron stars and the lightest observed black holes. Given the expected maximum mass values for neutron stars, this object was identified as a black hole, and, with the secondary component being a neutron star, the detection was classified as a neutron star-black hole merger. However, due to the low signal-to-noise ratio and the known waveform degeneracy between the spin and mass ratio in the employed gravitational-wave models, GW230529\_181500 could also be interpreted as a merger of two heavy ($\gtrsim 2 \mathrm{M}_\odot$) neutron stars with high spins. We investigate the distinguishability of these scenarios by performing parameter estimation on simulated signals obtained from numerical-relativity waveforms for both neutron star-black hole and binary neutron star systems, with parameters consistent with GW230529\_181500, and comparing them to the analysis of the real event data. We find that GW230529\_181500 is more likely to have originated from a neutron star-black hole merger, though the possibility of a binary neutron star origin can not be ruled out. Moreover, we use the simulation data to estimate the signatures of potential electromagnetic counterparts emitted by the systems. We find them to be too dim to be located by current wide-field surveys if only the dynamical ejecta is considered, and detectable by the Vera C. Rubin Observatory during the first two days after merger if one accounts for additional disk wind ejecta.
\end{abstract}

\maketitle

\section{\label{sec:level1}Introduction}

Since the first detection of gravitational waves (GWs) from a binary black hole (BBH) merger in 2015~\cite{LIGOScientific:2016aoc}, there have been numerous detections of compact binary coalescences (CBCs). One of those was the first detection of a binary neutron star (BNS) merger, GW170817, in 2017~\cite{LIGOScientific:2017vwq}. This GW detection together with the coincident gamma-ray burst (GRB) led to an unprecedented electromagnetic follow-up campaign, starting a new chapter in multi-messenger astronomy~\cite{LIGOScientific:2017ync,Goldstein:2017mmi,Coulter:2017wya,LIGOScientific:2017zic,LIGOScientific:2017adf,Kasen:2017sxr,Pian:2017gtc}.

Another kind of CBC, black hole-neutron star (BHNS) mergers, are also capable of ejecting matter but tend to have more neutron-rich and asymmetric ejecta~\cite{Kyutoku:2013wxa,Paschalidis:2014qra,Kiuchi:2015qua}. Similar to the BNS case, the expanding ejected matter undergoes $r$-process nucleosynthesis and subsequent radioactive decay, resulting in an electromagnetic transient lasting for a couple of days or weeks, also known as kilonova~\cite{Li:1998bw,Metzger:2010sy,Tanaka:2013ana}. The first BHNS merger observed via GWs was GW200115~\cite{LIGOScientific:2021qlt}. However, due to the parameters of the emitting system, the amount of ejecta was expected to be vanishingly small, making the corresponding electromagnetic counterpart extremely dim, if any. Indeed, no electromagnetic counterpart was found.~\cite{S200105aeCircs,2020GCN.26760....1K,2020GCN.26765....1U,2020GCN.26769....1U,2020GCN.26760....1K,2020GCN.26774....1G,2020GCN.26779....1S,2020GCN.26849....1R,2020GCN.26820....1N,2020GCN.26797....1S,Anand:2020eyg,Dichiara:2021vjy}.

On May 29th, 2023, GW230529\_181500 (hereafter GW230529) was observed by the LIGO Livingston GW detector~\cite{LIGOScientific:2024elc}. With signal-to-noise ratio (SNR) of around $11.3$ and a luminosity distance of $D_L = 201^{+102}_{-96}$~Mpc, GW230529 had an estimated source chirp mass of $\mathcal{M}_c=1.94^{+0.04}_{-0.04} \mathrm{M}_{\odot}$, a primary mass of $m_1= 3.6^{+0.8}_{-1.2} \mathrm{M}_\odot$ and a secondary mass $m_2= 1.4^{+0.6}_{-0.2} M_{\odot}$. This led to the primary interpretation of this event as a lower mass-gap object~\cite{Rhoades:1974fn,Kalogera:1996ci,Bailyn:1997xt,Ozel:2010su,Farr:2010tu}, presumably a black hole (BH), merging with a neutron star (NS) with a mass consistent with NS observations~\cite{Antoniadis:2016hxz,Ozel:2016oaf,Alsing:2017bbc,Farrow:2019xnc,Landry:2021hvl}. 
A mass-gap BH consistent with the primary component of GW230529 could form, e.g., as the product of a BNS merger~\cite{Mahapatra:2025agb}.
The follow-up campaigns did not find any electromagnetic counterparts for GW230529, primarily due to the expected low amount of ejecta and the large distance to the source~\cite{2023GCN.33890....1S,2023GCN.33894....1L,2023GCN.33892....1L,2023GCN.33900....1K,2023GCN.33980....1I,2023GCN.33897....1S,2023GCN.33893....1S,2023GCN.33896....1W,Ronchini:2024lvb,Pillas:2025pfc}.

However, the BHNS interpretation of this event is not the only one allowed within the statistical uncertainties of the detection. In principle, GW230529 can also be interpreted as a BNS merger of highly-spinning, heavy NSs with masses between $2.0$ and $2.4 \, \mathrm{M}_\odot$. For instance, in the original analysis of GW230529~\cite{LIGOScientific:2024elc}, the BNS interpretation becomes the most probable if a prior informed by a \textsc{Power Law + Dip + Break} population model ~\cite{Fishbach:2020ryj,Farah:2021qom,KAGRA:2021duu} is employed. The primary reason for this ambiguity in the interpretation of GW230529 as either BHNS or BNS is the low SNR of the event~\cite{Cotturone:2025jlm} coupled with the waveform degeneracy between the mass ratio and effective spin~\cite{Baird:2012cu}.
This degeneracy can be mitigated by employing astrophysically motivated priors, primarily due to the exclusion of large component spins~\cite{Chattopadhyay:2024hsf}.

In this paper, we explore the possibility that GW230529 originated from a BNS merger by performing general-relativistic hydrodynamics simulations of both BHNS and BNS systems with parameters consistent with GW230529. We assess the distinguishability of these scenarios by performing parameter estimation (PE) on the obtained numerical data and the real event data~\cite{GW230529data}.

The paper is structured as follows. In Sec.~\ref{section:degeneracy}, we discuss the existing degeneracy between the mass ratio and spin, which hinders the exact source classification of the GW230529 signal. We create a set of binary configurations that represent both interpretations and yield similar gravitational waveforms. Then, in Sec.~\ref{section:realistic_injections}, we present numerical-relativity (NR) simulations of these binary configurations evolved past merger, the extracted gravitational-wave signals, and ejecta profiles. We then hybridize these numerical waveforms with phenomenological waveform models, producing strain data that starts from the minimum detector frequency. In Sec.~\ref{section:parameter_estimation}, we perform PE on simulated signals obtained by injecting the hybrid waveforms in the detector noise estimated for GW230529.
We recover the binary parameters using different waveform models and compare the posteriors with those of GW230529. In Sec.~\ref {section:remnant_and_em}, we analyze the remnant properties, perform radiative transfer simulations on the numerical ejecta profiles to obtain kilonova lightcurves, and assess their observability. Finally, we discuss the results and conclude in Sec.~\ref{section:conclusion}. In Appendix~\ref{section:convergence}, we test the convergence of the numerical waveforms and, in Appendix~\ref{section:sampler_settings}, we investigate the influence of different sampler settings on the final posterior. Finally, in Appendix~\ref{section:carbon_footprint}, we estimate the energy used by this study and the induced greenhouse gas emissions.

Throughout the paper, we use geometrized units for all the physical quantities, where $G\mathbin{=}c\mathbin{=}M_\odot\mathbin{=}1$, unless stated otherwise.

\section{Probing the waveform degeneracy}
\label{section:degeneracy}

\subsection{Mass-spin degeneracy}

Gravitational signals from compact binaries are known to have a degeneracy between the mass ratio and the effective spin, see, e.g., Ref.~\cite{Baird:2012cu}. This degeneracy arises primarily from the spin-orbit phase contributions at the 1.5~post-Newtonian (PN) order in waveform expansions \cite{Kidder:1992fr,Poisson:1995ef,Cutler:1994ys}. 

In the aligned-spin case, the dominant spin contribution can be described by a single parameter, the reduced spin $\chi_{\mathrm{red}}$
\begin{equation}
    \chi_\mathrm{red} = \chi_\mathrm{eff} - \frac{38}{113} \eta (\chi_1 + \chi_2),
\end{equation}
where 
\begin{equation}
    \chi_\mathrm{eff}=\frac{\chi_1 m_1 + \chi_2 m_2}{M}
\end{equation}
is the effective spin, $\chi_{1,2}$ are the individual spin components, $M$ is the total mass, and $\eta=m_1 m_2 / M^2$ is the symmetric mass ratio. The phase evolution up to 1.5PN takes the following form~\cite{Poisson:1995ef,Ajith:2011ec}:
\begin{equation}
    \label{eq:phase_evolution}
    \begin{aligned}
    &\psi(v) = \frac{3}{128} \frac{1}{\eta v^5} \times \\  &\left( 1 + v^2 \left[ \frac{55}{9} \eta + \frac{3715}{756} \right] + v^3 \left[ \frac{113}{3} \chi_{\mathrm{red}} - 16 \pi \right] + \mathcal{O}(v^4) \right),
    \end{aligned}
\end{equation}

where $v=(M \omega_{\mathrm{orb}})^{1/3}$ is the invariant velocity, with $\omega_{\mathrm{orb}}$ being the orbital angular frequency of the binary.

Throughout most of the inspiral of a low-mass binary, $v$ changes only slightly\footnote{For the frequencies in 30-300\ Hz, $v$ changes only by around a factor of two~\cite{Baird:2012cu}.}. With such small changes in $v$, $\chi_{\mathrm{red}}$ and $\eta$ produce competing contributions at different PN orders to the phase evolution in Eq.~\eqref{eq:phase_evolution}, leading to a roughly linear correlation between the symmetric mass ratio $\eta$ and the reduced spin $\chi_{\mathrm{red}}$. Hence, systems with high spins can produce GW signals similar to the ones emitted by systems with more asymmetric mass ratios. The presence of this degeneracy complicates parameter estimation and hinders a clear source classification for GW230529.

\subsection{Selected GW230529-like binary configurations}
To explore the possibility of the BNS and BHNS interpretations of GW230529, we construct multiple configurations that are consistent with GW230529, i.e., their source chirp masses $\mathcal{M}_c$, mass ratios $q$, and effective spins $\chi_{\mathrm{eff}}$ lie within the 90\% credible interval of the GW230529 posteriors~\cite{LIGOScientific:2024elc}. Our configurations contain NSs with masses up to $2.4 \, M_\odot$.  For this reason, to describe NS matter in our systems, we select two equations of state (EOSs) that support such masses, the stiffer EOS DD2~\cite{Typel:2009sy} and the softer EOS MPA1~\cite{Read:2008iy}. Neither of these EOSs is excluded by the measurements of GW170817~\cite{LIGOScientific:2018cki}, but it is worth highlighting that many other state-of-the-art EOSs do not support such high NS masses. 

For the representative BHNS configuration, we pick the mass ratio and dimensionless effective spin to be $q=0.4$ and $\chi_{\mathrm{eff}} = -0.078$,  respectively, and we assume a non-spinning NS.

For the BNS systems, we set the component masses to be $m_1=2.40$ and $m_2=2.07$. The effective spin is $\chi_{\mathrm{eff}} = -0.28$. For this value of the effective spin, we pick two BNS spin configurations: BNS\_DS (double spinning), where both NSs have comparable negative spins, and BNS\_PS (primary spinning), where the primary NS is highly spinning and the secondary is non-spinning.

We summarize all configurations in Table~\ref{table:configurations}, along with their EOSs, tidal parameters, and other properties.

Additionally, we provide the rotational frequencies of the NSs calculated directly from our initial data. We note that while NSs with masses and spins of the ones in BNS\_DS configurations are known to exist (such as PSR~J0952-0607~\cite{Romani:2022jhd}), they are not found in BNS systems~\cite{Ozel:2016oaf}. Assembly of such systems should be extremely rare, as it would require two NSs to undergo long and independent recycling phases with their non-degenerate donor stars, and then dynamically rearrange from their natal tight NS-white dwarf binaries~\cite{Tauris:2012jp,Tauris:2017omb}.

\begin{table*}
    \centering
    \begin{tabular}{l|cccc|ccccc|cccccc|ccc}
        Name & $m_1$ & $m_2$ & $M$ & $\mathcal{M}_c$ & $\chi_1$ & $\chi_2$ & $\chi_{\mathrm{eff}}$ & $f_\mathrm{rot,1},$ & $f_\mathrm{rot,2},$ & EOS &   $\Lambda_1$ & $\Lambda_2$ & $\tilde{\Lambda}$ & $\mathcal{C}_1$ & $\mathcal{C}_2$ & $d$ & $M\Omega_{\mathrm{orb,0}},$ & $e,$  \\
         & &  &  &  & &  &  & [Hz] & [Hz] &  &  &  &  &  &  &  & $\mathrm{[10^{-2}]}$ & $\mathrm{[10^{-4}]}$  \\
        \hline
         \verb|BHNS_DD2| & $\BH$ 3.59  & 1.44 & 5.03 & 1.93982 & -0.11 & 0.0 & -0.078 & -- & 0 & DD2 &  0 & 595 & 43.5 & 0.5 & 0.161 & 58 & 2.29343 & $1$ \\
         \verb|BHNS_MPA1| & $\BH$ 3.59  & 1.44 & 5.03 & 1.93982 &  -0.11 & 0.0 & -0.078 & -- & 0 & MPA1 &  0 & 419  & 30.7 & 0.5 & 0.175 & 58  & 2.29109 & $8$ \\
         \verb|BNS_DD2_DS| & $\NS$ 2.40 & 2.07 & 4.47 & 1.93931 &  -0.3 & -0.25 & -0.277 & 768 & 545 & DD2 &  10.2 &  53.2  & 27.2 & 0.283 & 0.232 & 56 & 2.05536 & $4$ \\
         \verb|BNS_MPA1_DS| & $\NS$ 2.40 & 2.07 & 4.47  & 1.93931 &  -0.3 & -0.25  & -0.277 & 804 & 590  & MPA1 &  8.3 & 36.6  & 19.5 & 0.296 & 0.249 & 56 & 2.05670 & $2$ \\
         \verb|BNS_MPA1_PS| & $\NS$ 2.40 & 2.07 & 4.47  & 1.93931 &  -0.52 & 0.0 & -0.279 & 1240 & 0 & MPA1 &  10.3 & 35.3 & 20.2 &  0.289 & 0.250 & 56 & 2.05820 & $8$ \\
    \end{tabular}
    \caption{Binary parameters for the considered configurations. \textit{First group of columns:} $m_1$ is the gravitational mass of the primary object, which is, depending on the object type, the Christodolou mass of the BH or the ADM mass of the primary NS in isolation. $m_2$ is the gravitational mass of the secondary object, which is always an NS. $M$ is the total mass of the system, $\mathcal{M}_c$ is the chirp mass. \textit{Second group of columns:} $\chi_{\rm 1,2}$ are the dimensionless spin parameters, $\chi_{\mathrm{eff}}$ is the effective spin parameter, and $f_\mathrm{rot,1,2}$ are the NS rotational frequencies. \textit{Third group of columns:} EOS is the EOS employed, $\Lambda_{\rm 1,2}$ are the quadrupolar tidal deformabilities for the primary and secondary object, respectively, $\tilde{\Lambda}$ is the effective tidal deformability of the binary, $\mathcal{C}_{\mathrm{1,2}}$ are the compactnesses of each object, calculated for NSs by solving the TOV equation for a star with the same baryonic mass. \textit{Fourth group of columns:} $d$ is the initial binary separation, $M\Omega_{\mathrm{orb,0}}$ is the mass-scaled initial orbital frequency, and $e$ is the residual eccentricity in the initial data.}
    \label{table:configurations}
\end{table*}

\section{GW230529-like NR simulations}
\label{section:realistic_injections}

\subsection{Initial data}
To construct the initial data for our simulations, we use the \textsc{FUKA} initial data solver~\cite{Papenfort:2021hod}, which is based on the \textsc{Kadath} spectral library~\cite{Grandclement:2009ju}. \textsc{FUKA} solves the Einstein constraint equations using the extended conformal thin sandwich (XCTS) formalism~\cite{York:1998hy,Pfeiffer:2002iy}.

To obtain accurate waveforms, the residual eccentricity must be reduced to produce quasicircular binary inspirals. To achieve this, we use the eccentricity reduction procedure of Ref.~\cite{Papenfort:2021hod}\footnote{We perform the evolution of the system at low resolution (R128, cf.~Sec.~\ref{section:evolution}) and track the proper distance between the objects. For BNS systems, we use the proper distance along the coordinate line connecting the positions of the NS centers, and the proper distance from the NS center to the closest point to the NS on the apparent horizon surface for BHNS. For the model fitting, we use the time derivative of the proper distance. We terminate the procedure once the residual eccentricity $e$ is below $10^{-3}$.}. The final residual eccentricities $e$ for each configuration are quoted in Table~\ref{table:configurations}.

It was not possible to construct the initial data for one configuration, namely BNS\_DD2\_PS, which shares the parameters with BNS\_MPA\_PS except the EOS, due to a non-convergent solution in \textsc{FUKA}. The issue appears only when constructing a solution with the DD2 EOS and manifests in high-density oscillations at the NS surface. Together with the developers of \textsc{FUKA}, we perform an array of tests but are unable to resolve the issue.

\subsection{Evolution}
\label{section:evolution}
To evolve the binary configurations in time, we employ the general-relativistic hydrodynamics (GRHD) code \textsc{BAM}~\cite{Bruegmann:2006ulg,Thierfelder:2011yi,Dietrich:2015iva,Bernuzzi:2016pie,Dietrich:2018phi}. \textsc{BAM} uses the constraint-damping conformal Z4 (Z4c) reformulation \cite{Hilditch:2012fp} of Arnowitt-Deser-Misner (ADM) equations~\cite{Arnowitt:1962hi}, and the moving-puncture gauge \cite{Campanelli:2005dd,Baker:2005vv}. We employ high-resolution shock-capturing methods using WENO-Z reconstruction~\cite{2008JCoPh.227.3191B} with the Local Lax-Friedrichs (LLF) Riemann solver.

The grid in \textsc{BAM} consists of eight nested Cartesian levels. The two finest are allowed to move. The moving levels can contain separate subgrids (boxes) that cover each object and move dynamically with it. Each finer level has twice the spatial resolution of the corresponding parent level. In addition, the non-moving levels have twice as many points in one direction as the moving ones. Given this setup, the simulation resolutions are defined by the number of points in one direction on the moving levels, $n$, and thus, we name them as R$n$. Here, we run simulations at three resolutions: R96, R144, and R192. The finest levels fully cover the compact objects of the system, with the size of the level being $1.3$ times larger than the diameter of the NS. Hence, the grid setup depends on the stiffness of the EOS. In our BHNS cases, the BH is always fully located within a box at the finest level, and no additional refinement levels are added. We list the finest-level grid spacings for each resolution and configuration in Table~\ref{table:grid_spacings}.

\begin{table}[!htp]
    \centering
    \begin{tabular}{l|c|c|c}
        Configuration & $\Delta x^{\mathrm{R96}}~\mathrm{[m]}$ &  $\Delta x^{\mathrm{R144}}~\mathrm{[m]}$ &  $\Delta x^{\mathrm{R192}}~\mathrm{[m]}$\\
        \hline
        \verb|BHNS_DD2| & 272 & 181 & 136 \\
        \verb|BHNS_MPA1| & 247 & 165 & 124 \\
        \verb|BNS_DD2_DS| & 274 & 182 & 137 \\
        \verb|BNS_MPA1_DS| & 249 & 166 & 125 \\
        \verb|BNS_MPA1_PS| & 255 & 170 & 127 \\
        \end{tabular}
    \caption{Finest level grid spacings at three resolutions: R96, R144, and R192.}
    \label{table:grid_spacings}
\end{table}

\begin{figure}[!htp]
    \centering
    \includegraphics[width=1\linewidth]{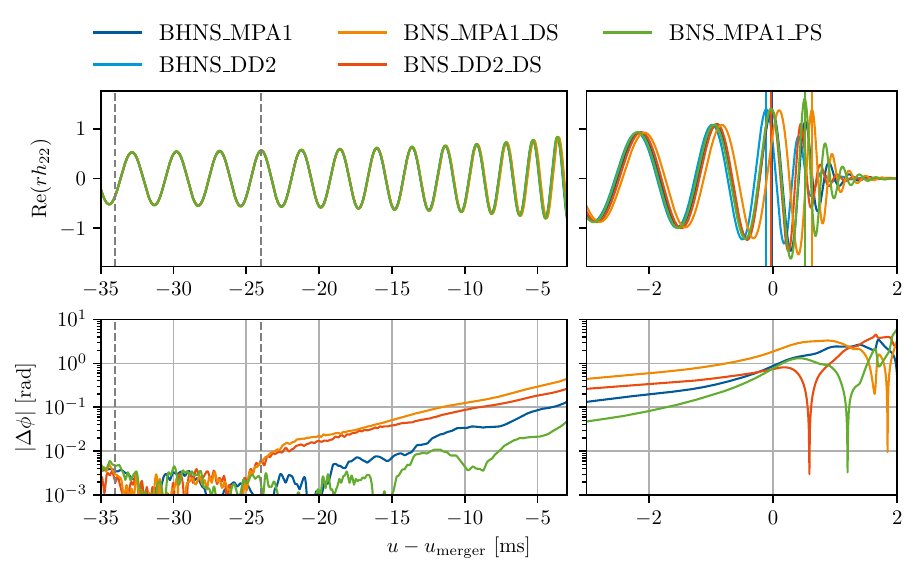}
    \caption{\textit{Top panels:} Real part of the (2,2)-mode of the spin-weighted spherical harmonic coefficients as a function of retarded time before merger with the extraction radius $r=1200$ scaled out. The merger time reference is taken to be that of the BHNS\_DD2 configuration. The vertical solid lines show the corresponding merger times for each configuration. The vertical dashed gray lines indicate the alignment window. \textit{Bottom panel:} Phase differences between each configuration with respect to the BHNS\_DD2 one.}
    \label{figure:nr_waveforms_diff}
\end{figure}

In Figure~\ref{figure:nr_waveforms_diff}, we show the numerical waveforms at the highest resolution, R192, and aligned between 34 and 24~ms before merger. The waveforms agree up to a total dephasing at merger of about 1~rad accumulated throughout the simulation.  
Both BHNS cases and BNS\_DD2\_DS have almost coincident merger times, while BNS\_MPA1\_DS and BNS\_MPA1\_PS merge noticeably later. We attribute it to the lower compactness and higher tidal deformability of the DD2 stars, which leads to an earlier merger. The robustness of the numerical waveforms is verified in a detailed convergence study in Appendix~\ref{section:convergence}.

\subsection{Hybridization with waveform models}
To reduce the effects of the finite-radius extraction, we perform a perturbative extrapolation of the waveforms using the \textsc{scri} software package~\cite{mike_boyle_2020_4041972,Boyle:2013nka,Boyle:2014ioa,Boyle:2015nqa}. Here, we use the waveforms with the extrapolation order $N=2$, as for the higher extrapolation orders, the numerical noise is significantly amplified.

\begin{figure}[htp]
    \centering
    \includegraphics[width=1\linewidth,height=\textheight,keepaspectratio]{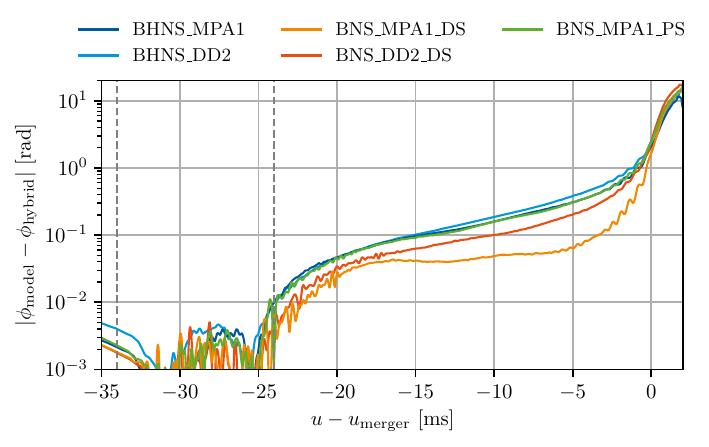}
    \caption{Phase difference between the waveform models and the hybrids for all the configurations. The vertical dashed gray lines signify the alignment window.} 
    \label{figure:hybrids_vs_models}
\end{figure}

The NR waveforms obtained from the simulations contain only the last $\sim18$ cycles. To perform an injection, a waveform spanning the entire frequency band of the detector must be created, as the detector is most sensitive to the inspiral part of the GW230529 signal. For that purpose, we hybridize the NR data with the waveform approximant \verb|IMRPhenomXAS_NRTidalv3|~\cite{Pratten:2020fqn,Abac:2023ujg}. For the hybridization, the numerical waveforms are aligned by minimizing the phase difference integral
\begin{equation}
    I = \int_{t_i}^{t_f} |\phi_A(t+\delta t) + \delta \phi - \phi_B|^2 dt,
\end{equation}
where $t_i$ and $t_f$ are the start and end times of the alignment window, $\phi_A$ is the phase of the waveform to be aligned, $\phi_B$ is the phase of the waveform to which $\phi_A$ is aligned, and $\delta t$ and $\delta \phi$ are the time and phase adjustments, respectively. 
We use as start and end times 34~ms and 24~ms before the merger, respectively.  
The waveforms are then hybridized using the Tukey blending function defined in Eq.~(23)~of Ref.~\cite{MacDonald:2011ne}. The phase differences between the hybrids and the \verb|IMRPhenomXAS_NRTidalv3| waveforms are shown in Figure~\ref{figure:hybrids_vs_models}.

The hybrid waveforms are then projected onto the LIGO Livingston detector frame, assuming distance and sky location corresponding to the maximum likelihood values obtained with the PE analysis in the GW230529 detection paper~\cite{LIGOScientific:2024elc}. We consider only the LIGO Livingston detector because it was the only one that observed GW230529. The data are then injected into detector noise generated from the power spectral density estimated for GW230529 between 20 and 2048 Hz. Both of these steps are performed using the \textsc{PyCBC} package~\cite{pycbc}.

\section{Parameter estimation}
\label{section:parameter_estimation}
For our PE runs, we employ the Bayesian inference framework \textsc{bilby}~\cite{Ashton:2018jfp} with the dynamic nested sampler \textsc{dynesty}~\cite{Speagle:2019ivv,sergey_koposov_2024_12537467}. For recovery, we select a set of phenomenological GW models with varying assumptions and modeling complexity: aligned spins, precession, higher-order modes, and tides. We list the models with their corresponding references and main features in Table~\ref{tab:gw_models}. All of the selected waveforms are implemented in \textsc{LALsuite}~\cite{lalsuite}. 
For our injections, we use only the dominant (2,2) mode. To justify this approach, we have tested waveform-model injections incorporating higher modes for both BHNS and BNS configurations and found nearly identical posteriors when including and excluding higher-order modes.

\begin{table}[!htbp]
    \centering
    \begin{tabular}{l|c|c|c|c}
       Model                & References & Precession & HM & Tides \\
       \hline
       \verb|IMRPhenomXAS| &   \cite{Pratten:2020fqn}     &      -     &  -   &   - \\
       \verb|IMRPhenomXP| &      \cite{Pratten:2020ceb}     & \cmark         &   -  &   - \\
       \verb|IMRPhenomXHM| &    \cite{Garcia-Quiros:2020qpx}       &     -      & \cmark   &   -  \\
       \verb|IMRPhenomXPHM| &      \cite{Pratten:2020ceb}     & \cmark         & \cmark   &  -  \\
       \verb|IMRPhenomXAS_NRTidalv3| &    \cite{Pratten:2020fqn,Abac:2023ujg}     &       -    &  -  & \cmark  \\
       \verb|IMRPhenomXP_NRTidalv3| &  \cite{Pratten:2020ceb,Abac:2023ujg}         & \cmark         &  -   & \cmark  \\
       \verb|IMRPhenomNSBH| &   \cite{Thompson:2020nei}        &    -       &   -  & \cmark  \\
    \end{tabular}
    \caption{The gravitational-wave models used in this work. The columns represent the model name in \textsc{LALSuite}, the corresponding references, and whether the model includes precession, higher modes (HM), and tides, respectively.}
    \label{tab:gw_models}
\end{table}

We perform PE on the injection frames using sampler settings close to the PE runs performed in the GW230529 discovery paper~\cite{LIGOScientific:2024elc}, with additional adjustments. We use nested sampling with 2000 live points, differential evolution~\cite{Braak2006} for the live point proposal, a \textsc{Bilby}-specific acceptance walk-stepping method with an average of 60 accepted chains, and a stricter stopping criterion with the threshold for the estimated remaining contribution to the log-evidence at $d\log{\mathcal{Z}}=10^{-3}$. A detailed comparison of the impact of different sampler settings that drove this choice can be found in Appendix~\ref{section:sampler_settings}. To accelerate the likelihood computation, we employ a multibanding technique~\cite{Morisaki:2021ngj}. 
As the systematic error of the detector calibration is expected to be subdominant to the systematics of the waveform models~\cite{Payne:2020myg}, we do not include it in our analysis.

\subsection{Choice of priors}
For the analysis of the injected signals, we use uniform priors in chirp mass $\mathcal{M}_c \in [1.935, 1.945]$, i.e., a $0.01$-wide range centered at the injected value\footnote{The chirp mass for the injections is not redshifted as the waveform was not rescaled for a specific distance.}, and $\mathcal{M}_c \in [2.0214, 2.0331]$ for the analysis of the GW230529 data. 
We use a uniform prior in mass ratio $q \in [0.2, 1.0]$. 

We consider wide priors in component spins, i.e. $|\chi_{1,2}| \in [0, 0.99]$, consistent with the GW230529 analysis\footnote{Applying this wide spin prior $|\chi_{1,2}| < 0.99$ to IMRPhenomNSBH is not possible due to the limitations of the fitting formula for the final BH spin~\cite{Rezzolla:2007rd} used internally via IMRPhenomC routines~\cite{Santamaria:2010yb}. For large effective spins and more unequal mass ratios, the fitting formula predicts remnant spins higher than the Kerr limit. For the lowest mass ratio in the prior, $q=0.2$, the maximum allowed component spin is $\chi_{1,2}=0.96$.  Because of this limitation, we reduce the maximum of the spin magnitude prior for IMRPhenomNSBH to that value, and still include it in the comparisons with this minor caveat.}. For the waveform models with aligned spins, we choose two kinds of priors, the \texttt{AlignedSpin} (AS) and \texttt{DiscreteValues} (DV) priors.
The AS prior is the standard \textsc{Bilby} prior for aligned-spin approximants and was also used in the original analysis of the event. It is constructed by the projection of spins uniformly distributed in both magnitude and direction on the direction of the binary's orbital angular momentum~\cite{Ashton:2018jfp,Lange:2018pyp}.
Conversely, the DV prior is constructed to have uniformly distributed spin magnitude in the direction of the binary's orbital angular momentum and uniformly chosen discrete spin signs, i.e., aligned or anti-aligned. Compared to the AS prior, which has most of the effective spin prior mass around zero, the resulting effective spin prior probability for the DV case is significantly flatter and closer to uniform. This yields more support for high spin values compared to the AS prior.

For the precessing models, we use the spin prior where the component spins are isotropic in direction and uniform in magnitude.
For the models with tides, we employ uniform priors in the individual tidal deformabilities $\Lambda_{1,2} \in [0, 5000]$. We marginalize the likelihood over the luminosity distance to the source $D_L$, with a prior uniform in comoving volume $D_L \in [10, 500]$~Mpc. We use the LIGO Livingston detector's time as the time reference instead of the geocentric time -- as GW230529 was a single-detector event, this removes the ambiguity of the detector's position relative to the Earth's center. For the other extrinsic parameters of the signal, such as the polarization angle and coalescence phase, we use the standard priors as in Ref.~\cite{LIGOScientific:2024elc}.

\begin{figure*}[t]
    \centering
    \includegraphics[width=0.48\linewidth]{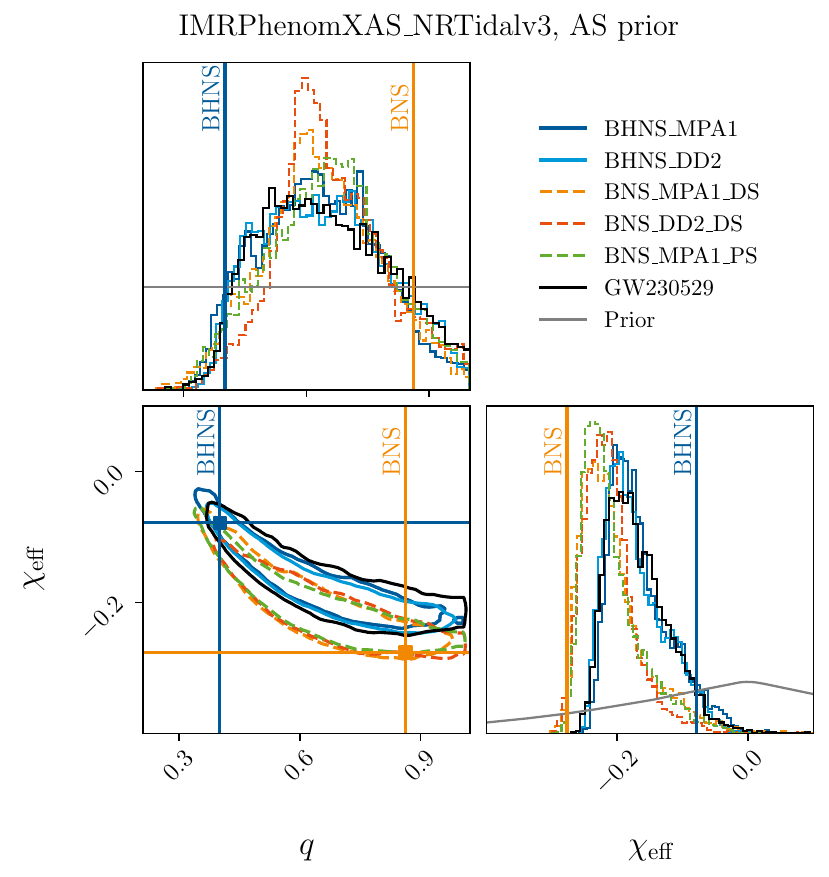}
    \includegraphics[width=0.48\linewidth]{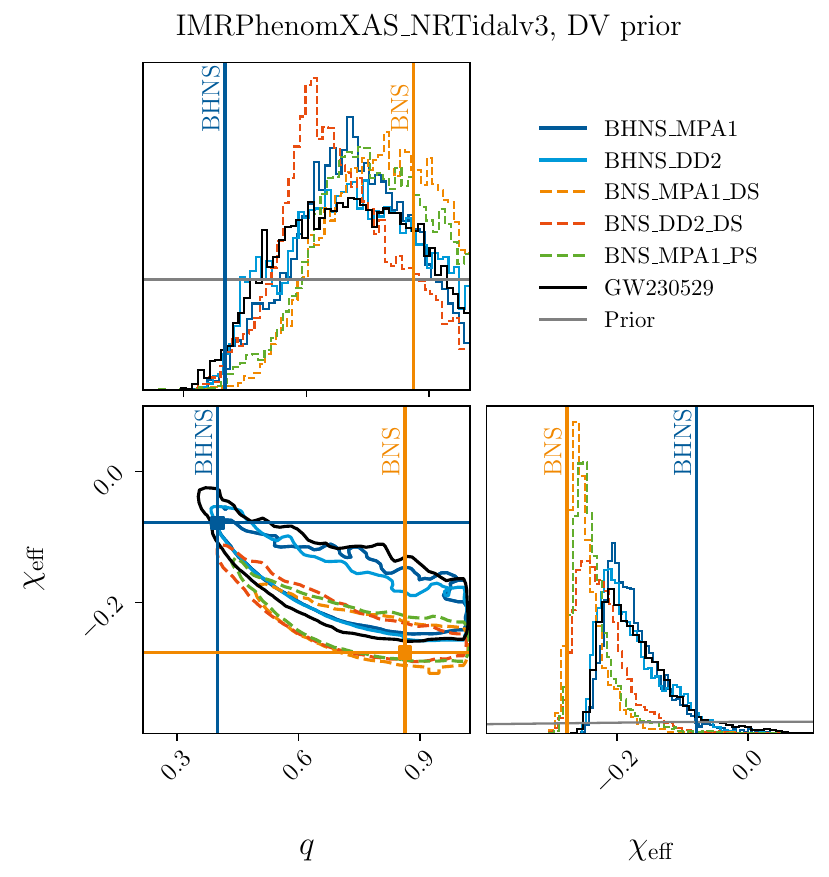}
    \caption{Posterior distributions for the mass ratio $q$ and effective spin $\chi_{\mathrm{eff}}$ for the analysis with \texttt{IMRPhenomXAS\_NRTidalv3} and the AS prior (\textit{left}) and the DV prior (\textit{right}) for the injections and for the GW230529 event data. The contours in the $q$-$\chi_{\mathrm{eff}}$ panel correspond to the 90\% credible level. The orange vertical and horizontal lines mark the BHNS injected parameters, while the blue ones mark the BNS ones. The priors for both quantities are shown as gray lines. To aid the visual comparison, the prior probability densities for the effective spin are multiplied by four.} 
\label{figure:multicorner_v4_IMRPhenomXAS_NRTidalv3}
\end{figure*}

\subsection{Recovery using the injection waveform model}
\label{section:pe_injection_model}
Since \verb|IMRPhenomXAS_NRTidalv3| was used for the hybridization, we focus first on the results obtained using the same waveform model for recovery. In what follows, we focus on the mass ratio and the effective spin, as those are both astrophysically relevant and well-constrained intrinsic parameters of binaries.

In Figure~\ref{figure:multicorner_v4_IMRPhenomXAS_NRTidalv3}, we show the posterior distributions of $q$ and $\chi_{\mathrm{eff}}$ for all the injected hybrids and the GW230529 event data recovered using both the AS and DV priors. In the analysis, we are unable to narrowly recover the injections, and the recovered values lie close to the boundary of the 90\% credible interval. The posteriors for the mass ratio agree with each other regardless of the injection, and peak at values between the injected ones, especially in the case of the AS prior. For the DV prior, however, there is a mild skew of the mass ratio posteriors towards the BNS injected values.

The main difference between the BHNS and BNS cases lies in the $\chi_{\mathrm{eff}}$ posteriors. For BNS injections, the $\chi_{\mathrm{eff}}$ posteriors are slightly skewed towards less negative values with respect to the injected ones, while the skew is stronger and towards more negative values for BHNS, and is present for both injection types and regardless of the employed prior. This skew can be explained by the low SNR of GW230529, which additionally allows mutual compensation of more negative effective spin measurements with the lower chirp mass measurements. 
For the DV prior, the $\chi_{\mathrm{eff}}$ posteriors for the BNS cases are more peaked than those for the AS prior, primarily due to larger prior mass at more negative values of effective spin under the DV prior.

Finally, the $\chi_{\mathrm{eff}}$ posteriors for GW230529 are largely overlapping with the ones for the BHNS injections for both AS and DV priors.

\subsection{Recovery using waveform models with aligned spins}
\label{section:pe_aligned_spin}
\begin{figure*}[htp]
    \centering
    \includegraphics[width=0.48\linewidth,height=\textheight,keepaspectratio]{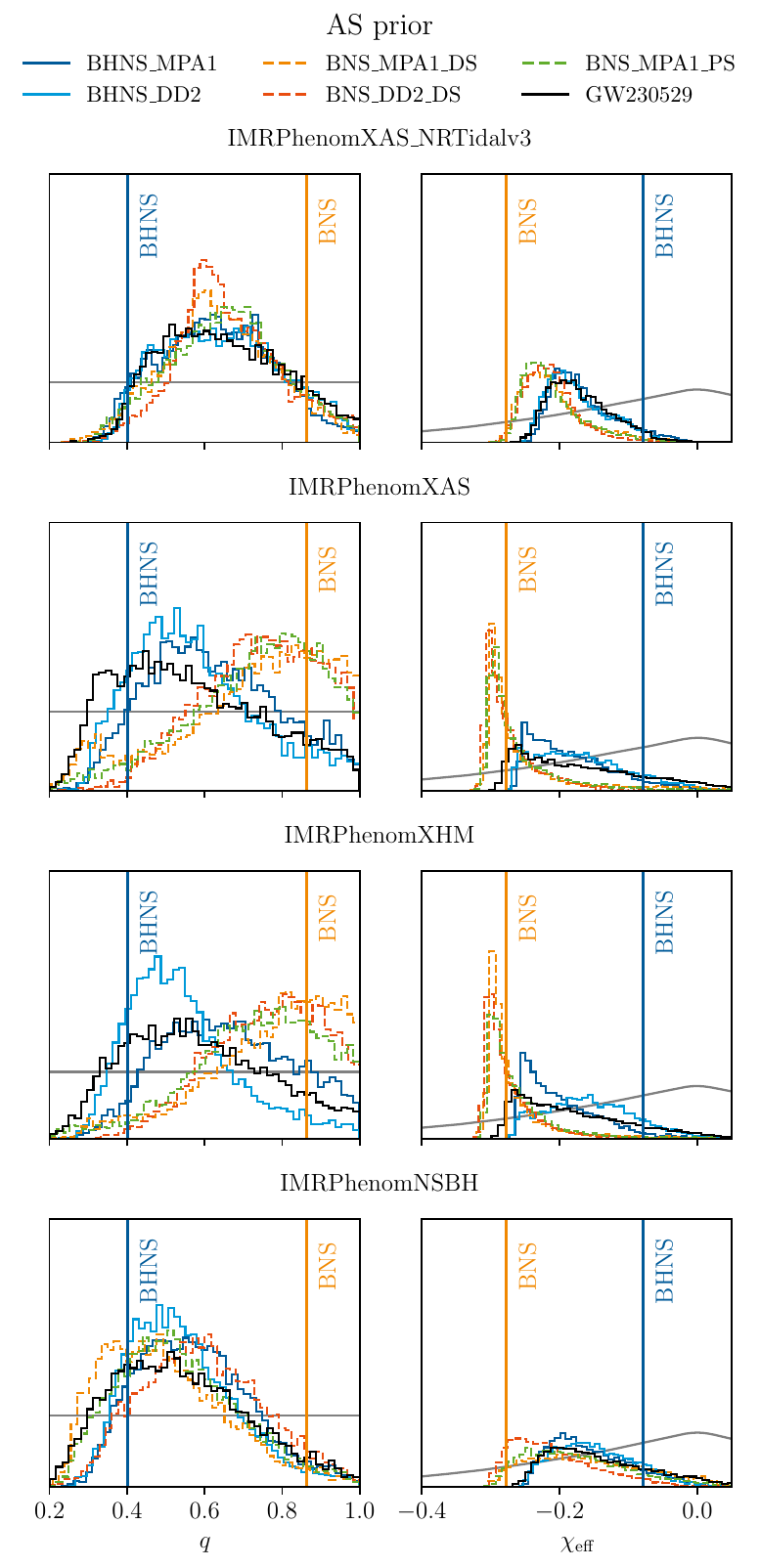}
    \includegraphics[width=0.48\linewidth,height=\textheight,keepaspectratio]{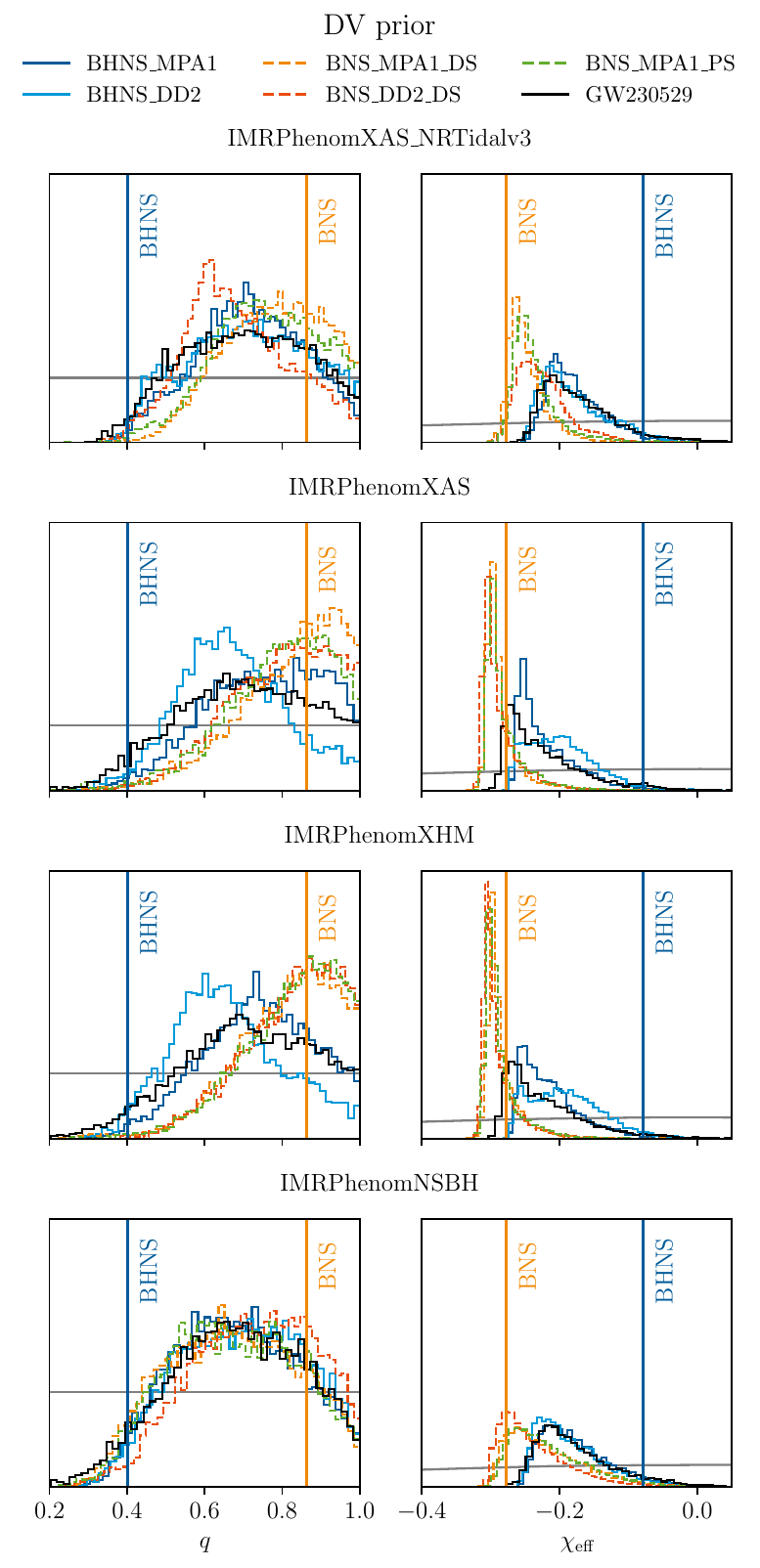}
    \caption{Posterior distributions for $q$ and $\chi_{\mathrm{eff}}$ recovered with different waveform models using the AS prior (\textit{left two columns}) and the DV prior (\textit{right two columns}). The vertical lines mark the injected values for the BHNS and BNS cases. The priors for both quantities are plotted as grey lines. To aid the visual comparison, the prior probability densities for the effective spin are multiplied by four.}
    \label{figure:qchi_v4_aligned_spins}
\end{figure*}

In Figure~\ref{figure:qchi_v4_aligned_spins}, we show posterior distributions for $q$ and $\chi_{\mathrm{eff}}$ recovered with different aligned-spin waveform models under both AS and DV priors. All of the models exhibit the following trends in the posterior distributions: The effective spins are not well-recovered, especially for the BHNS injections. The effective spin posteriors in the case of the DV prior have slightly more pronounced peaks, which is primarily due to the larger prior mass for the more negative values than in the case of the AS prior. The $\chi_{\mathrm{eff}}$ posteriors for the BHNS and BNS injections display clustering of overlapping distributions, thus giving a hint for a possible distinction between these two types of injections.

\verb|IMRPhenomXAS| is able to recover the injected mass ratios for all injections under the AS prior, but only for the BNS cases under the DV prior. For the DV prior, the mass ratio posteriors of the BHNS injections are shifted towards the BNS values. The $\chi_{\mathrm{eff}}$  posteriors have sharper peaks than for \verb|IMRPhenomXAS_NRTidalv3| in the case of BNS, and wider for BHNS. 

\verb|IMRPhenomXHM| produces posteriors similar to the \verb|IMRPhenomXAS| case. This is expected as the injected signal did not include higher modes. In both cases, the main distinctive feature is the more pronounced peak in the $q$ posterior for BHNS\_DD2 injection using the AS prior. 

\verb|IMRPhenomNSBH| results in similar posteriors for the mass ratio, without any strong dependence on the type of the injection. The $\chi_{\mathrm{eff}}$ posteriors, however, show the same distinction between BHNS and BNS injections. Compared to other models, both BHNS and BNS posteriors are less peaked and have extended tails. 

We observe that the aligned-spin BBH waveform models recover the mass ratio overall better than models with tides,  \verb|IMRPhenomXAS_NRTidalv3| and \verb|IMRPhenomNSBH|, which we attribute to the increase in the dimensionality of the parameter space by including tidal parameters and their correlation with the mass ratio. The latter, in turn, also affects the posteriors of the effective spin.

Irrespective of the waveform model, in the case of aligned spins, the GW230529 posteriors largely overlap with those of the BHNS injections.

\subsection{Recovery using waveform models with precession}
\label{section:pe_precession}
\begin{figure}[!htp]
    \centering
    \includegraphics[width=1\linewidth,height=\textheight,keepaspectratio]{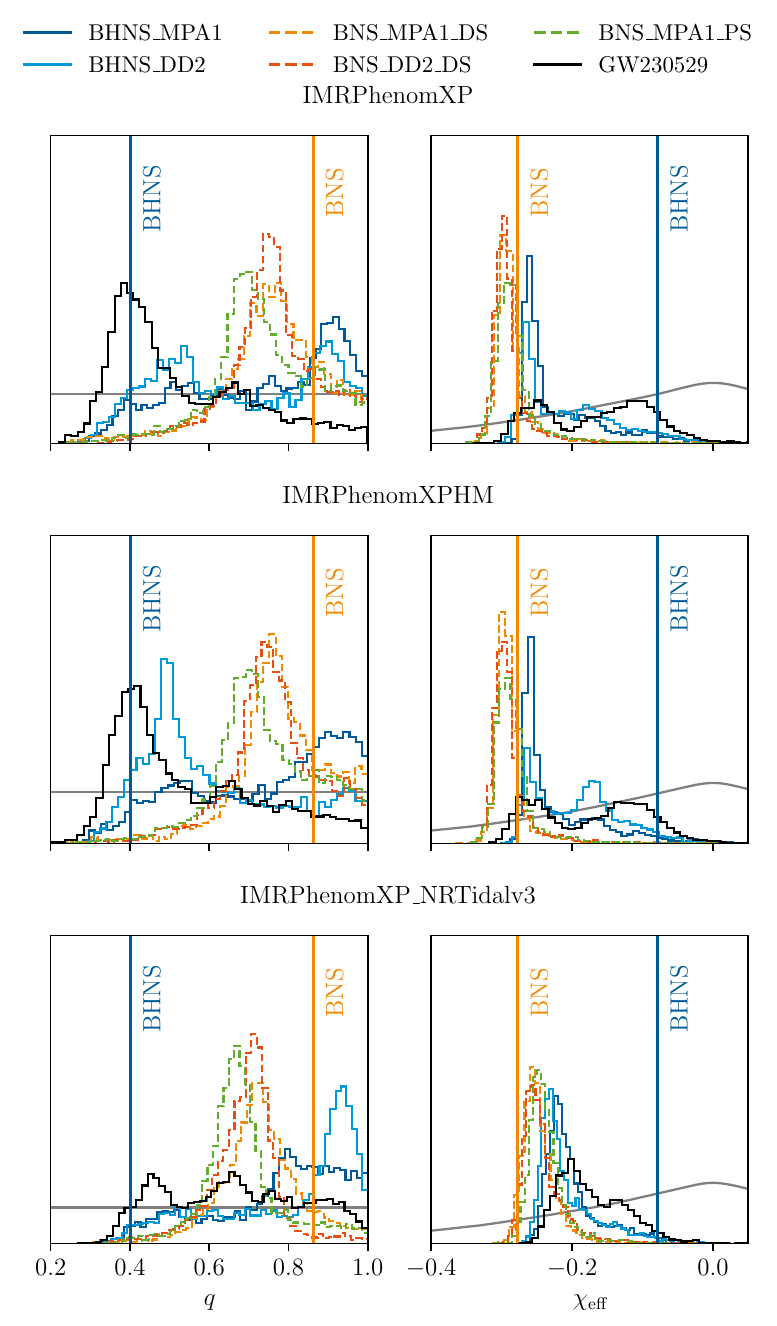}
    \caption{Same as for Figure~\ref{figure:qchi_v4_aligned_spins} but for precessing waveforms.} 
    \label{figure:qchi_v4_precessing}
\end{figure}

The results for the precessing waveform models are shown in Figure~\ref{figure:qchi_v4_precessing}. Compared to the posteriors for the models with aligned spins, the posteriors for the models with precession exhibit a more diverse behavior.

For the BNS injections, the mass ratio posteriors generally agree across the waveform models used for recovery and peak at around $0.7$, a value lower than the injected one. For \verb|IMRPhenomXP| and \verb|IMRPhenomXPHM|, the $q$ and $\chi_{\mathrm{eff}}$ posteriors for the BNS injections are largely similar to each other.

For the BHNS injections, the posteriors for the mass ratio are wider and have multimodal peaks at vastly different values. For the precessing waveforms without tides, \verb|IMRPhenomXP| and \verb|IMRPhenomXPHM|, the BHNS\_MPA1 mass ratio posterior peaks at higher values, close to that of the BNS injection. On the other hand, the mass ratio posteriors for the BHNS\_DD2 injection peak close to the injected BHNS value. For these models, the GW230529 posteriors peak sharply at the injected BHNS value. For the precessing waveform with tides, \verb|IMRPhenomXP_NRTidalv3|, the BHNS and GW230529 posteriors in mass ratio are significantly wider and flatter than for other precessing waveform models. The BHNS injections posteriors for $q$ have a slight preference for higher mass ratios.

As before, the main difference between the injections appears in the measurement of the effective spin -- the posteriors for BHNS and BNS form distinctive groups of similar distributions.ad The BHNS cases peak at more negative values away from the injected value and have more extended tails up to $\chi_{\mathrm{eff}} = 0$.
 The GW230529 $\chi_{\mathrm{eff}}$ posteriors are largely mimicking the behavior of the posteriors for the BHNS injections, though they have extended tails closer to the injected BHNS value.
 In the case of \verb|IMRPhenomXP_NRTidalv3|, the $\chi_{\mathrm{eff}}$ posteriors have less pronounced peaks, although they are overall narrower, especially in the case of BHNS injections.
Similarly to the original analysis, we find the posteriors for the effective precession spin parameter $\chi_p$~\cite{Schmidt:2014iyl} rather uninformative and prior-driven. 
However, due to the inclusion of extra degrees of freedom in the processing models, the posteriors slightly change in comparison to the aligned-spin scenario, leading, in some cases, to a narrower posterior for the mass ratio and effective spin~\cite{Chatziioannou:2014coa}.

\subsection{Posterior similarity}
\label{section:posterior_similarity}
In order to quantify the similarity of GW230529 with the injections, we compare their posteriors using the Jensen-Shannon divergence (JSD)\footnote{To fully assign the probabilities to each scenario, one would need to compute the posterior odds by multiplying the Bayes factor by the prior odds. The prescription of the latter requires substantial population modeling, which is out of scope of this paper, as we focus on the distinguishability of the signals solely in PE.}~\cite{Lin:1991zzm}. For two probability distributions $P_1$ and $P_2$, the JSD is defined as
\begin{equation}
    \mathrm{JSD}(P_1, P_2) = \sqrt{\frac{1}{2} \left[ \mathrm{KL}(P_1, M) +     \mathrm{KL}(P_2,M) \right]},
\end{equation}
where KL is the Kullback-Leibler divergence~\cite{Kullback:1951zyt},
\begin{equation}
    \mathrm{KL}(P, M) = \sum_x P(x) \log{\frac{P(x)}{M(x)}}
\end{equation}
and $M = \frac{1}{2} \left[ P_1 + P_2 \right]$ is the mixture distribution between $P_1$ and $P_2$. Thus, for increasingly similar probability distributions, the JSD is approaching zero and vanishes for identical distributions.

\begin{figure*}[htp]
    \centering
    \includegraphics[width=0.48\linewidth,height=\textheight,keepaspectratio]{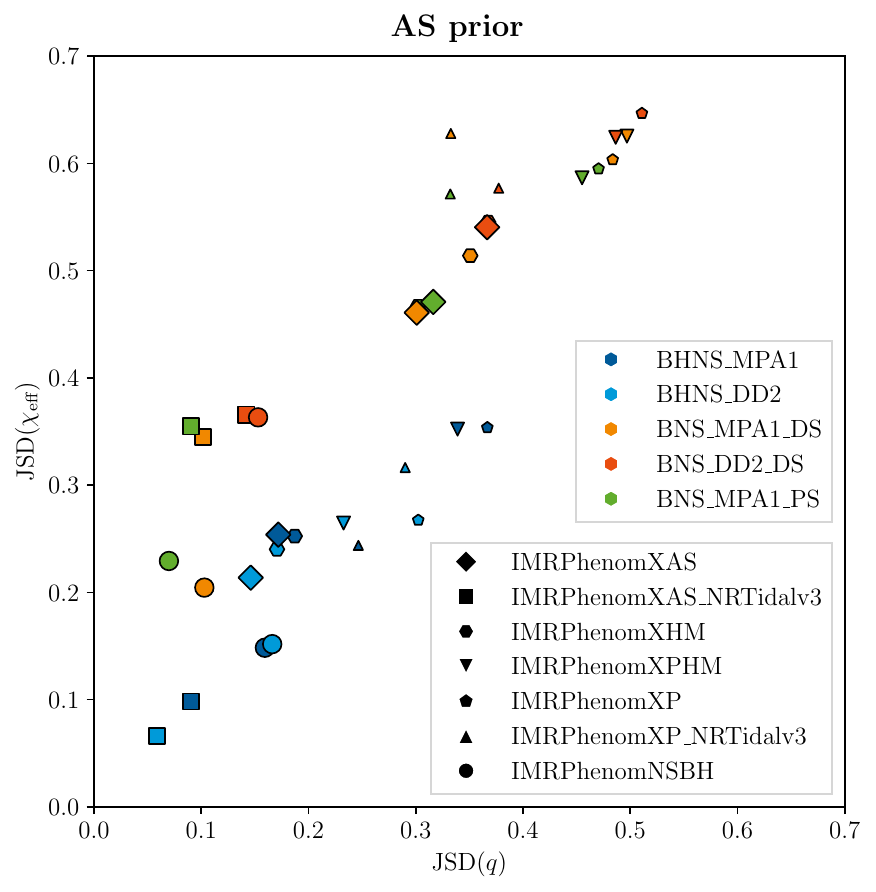}
    \includegraphics[width=0.48\linewidth,height=\textheight,keepaspectratio]{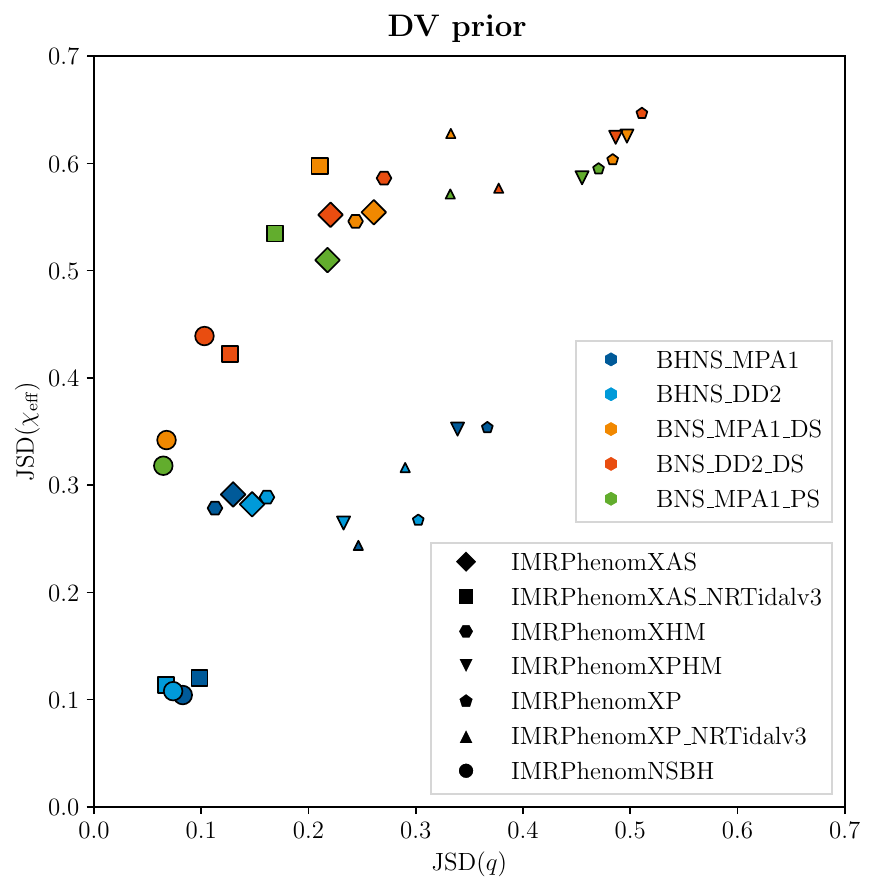}
    \caption{Jensen-Shannon divergence between the posterior distributions of GW230529 and the injections for $q$ and $\chi_{\mathrm{eff}}$ using the AS prior (\textit{left}) and the DV prior (\textit{right}). The data points for precessing waveforms are the same on both panels. Each color represents a different injection, and each marker represents the waveform model used for PE. The marker size is made unique for each waveform model to aid visual grouping.}
    \label{figure:jsd_clusters_v4}
\end{figure*}

We calculate the JSD between the posteriors of GW230529 and of all injections for $q$ and $\chi_{\mathrm{eff}}$ separately. The results are shown in Figure~\ref{figure:jsd_clusters_v4}.

Both the BHNS and BNS posteriors cluster according to their source type, with the BHNS cases being more similar to the GW230529 posteriors in the vast majority of cases and regardless of the spin prior. As expected from the analysis presented in the previous subsection, the BHNS cases have a lower JSD for the effective spin than their BNS counterparts.

Under the DV prior, the \verb|IMRPhenomXAS_NRTidalv3| and \verb|IMRPhenomNSBH| posteriors for the BHNS injections have more similar JSD values for both quantities compared to the ones obtained under the AS prior. For the same models, the $\chi_{\mathrm{eff}}$  posteriors for the BNS injections become more different from the ones of GW230529. Notably, the posteriors obtained with these models are distinctly closer to the GW230529 ones compared to the analysis with other approximants.
For \verb|IMRPhenomXAS| and \verb|IMRPhenomXHM|, the $q$ posteriors become generally more similar, and the differences in the $\chi_{\rm eff}$ ones attenuate.

Based on the posterior similarity, we find that GW230529 was more likely emitted by a BHNS merger than a BNS merger. However, the BNS merger interpretation cannot be ruled out.

\section{Remnant properties and electromagnetic signatures}
\label{section:remnant_and_em}
Beyond GW signals, NR simulations model the remnant of the mergers and thus allow calculation of their properties. We first summarize various remnant properties in Table~\ref{table:remnant_properties}, and discuss them separately.

\begin{table}[!bp]
    \centering
    \begin{tabular}{l|r|r|cc}
        Name & $M_{\mathrm{ej}}^{\mathrm{dyn}}$ & $M_{\mathrm{b}}^{\mathrm{rem}}$  & $M_{\mathrm{BH}}^{\mathrm{rem}}$  &  $\chi_{\mathrm{BH}}^{\mathrm{rem}}$ \\
        \hline
        \verb|BHNS_DD2| & $ 1.1 \times 10^{-3}$ &  $3 \times 10^{-2}$ & 4.895 & 0.568 \\
        \verb|BHNS_MPA1| & $ 1.2 \times 10^{-4}$ & $7 \times 10^{-3}$ & 4.901 & 0.566 \\
        \verb|BNS_MPA1_DS| & $5 \times 10^{-6}$ & $\lesssim 2 \times 10^{-5}$& 4.336 & 0.643 \\
        \verb|BNS_DD2_DS| & $2 \times 10^{-6}$ & $\lesssim 1 \times 10^{-4}$ & 4.347 & 0.657 \\
        \verb|BNS_MPA1_PS| &  $\lesssim 5 \times 10^{-8}$ & $\lesssim 1 \times 10^{-7}$ & 4.318 & 0.608 \\
         
    \end{tabular}
    \caption{Dynamical ejecta mass  $M_{\mathrm{ej}}^{\mathrm{dyn}}$, baryonic mass of the remnant $M_{\mathrm{b}}^{\mathrm{rem}}$, remnant BH mass $M_{\mathrm{BH}}^{\mathrm{rem}}$ and spin $\chi_{\mathrm{BH}}^{\mathrm{rem}}$ for the binary configurations.}
    \label{table:remnant_properties}
\end{table}

\subsection{Remnant BH}
Within each merger type, BNS or BHNS, remnant BHs have roughly similar masses. Slightly more massive remnant BHs are produced for BHNS configurations, which are mostly driven by their higher total mass. The BHNS cases with the stiffer DD2 EOS have higher disk mass and less massive remnant BH, in agreement with GW230529-like BHNS simulations~\cite{Matur:2024nwi}. For the BNS cases with MPA1 EOS, both the spin and the remnant mass are higher for the configurations where both NSs are spinning. This can be explained by the larger total angular momentum of the BNS\_MPA1\_DS system.

\begin{table}[!htbp]
    \centering
    \begin{tabular}{l|cc|cc|c}
        Name & $M_{\mathrm{BH}}^{\mathrm{rem}}$ & $\Delta M_{\mathrm{BH}}^{\mathrm{rem}}$  & $\chi_{\mathrm{BH}}^{\mathrm{rem}}$  &  $\Delta \chi_{\mathrm{BH}}^{\mathrm{rem}}$ & Reference \\
        \hline
        \verb|BHNS_DD2| & 4.885 &   $- 0.2\%$  & 0.565 & -0.5\% & \cite{Gonzalez:2022prs} \\
        \verb|BHNS_MPA1| & 4.904 &   $- 0.6\%$  & 0.566 & $\lesssim$ 0.1\% & \cite{Gonzalez:2022prs} \\
        \hline
        \verb|BHNS_DD2| & 4.893 &   $- 0.04\%$  & 0.564 & -0.7\% & \cite{Gonzalez:2025xba} \\
        \verb|BHNS_MPA1| & 4.907 &   $- 0.2\%$  & 0.564 & -0.4\% & \cite{Gonzalez:2025xba} \\
        \hline
        \verb|BNS_DD2_DS| & 4.327 &   $-0.46\%$  & 0.80 & 22\% & \cite{Coughlin:2018fis} \\
        \verb|BNS_MPA1_DS| & 4.329 &   $-0.17\%$  & 0.80 & -25\% & \cite{Coughlin:2018fis} \\
        \verb|BNS_MPA1_PS| & 4.328 &   $0.24\%$  & 0.79 & 30\% & \cite{Coughlin:2018fis} \\
             \end{tabular}
    \caption{Remnant mass $M_{\mathrm{BH}}^{\mathrm{rem}}$ and spin $\chi_{\mathrm{BH}}^{\mathrm{rem}}$ predicted by different analytical fitting formulae. $\Delta M_{\mathrm{BH}}^{\mathrm{rem}}$ and $\Delta \chi_{\mathrm{BH}}^{\mathrm{rem}}$ represent the relative difference to our NR data.}
    \label{table:remnant_properties_fits}
\end{table}

We compare the mass and spin of the remnant BH of our simulations to the predictions of the fitting formula of Ref.~\cite{Gonzalez:2022prs}, which was calibrated on a wide range of BHNS simulations, including low-mass and equal-mass ones. The results are presented in Table~\ref{table:remnant_properties_fits}. The formula is in excellent agreement with our simulation data -- it yields values for $M_{\mathrm{BH}}^{\mathrm{rem}}$ and $\chi_{\mathrm{BH}}^{\mathrm{rem}}$ that differ from the NR data by less than 1~\%. Employing the updated formula of Ref.~\cite{Gonzalez:2025xba} yields noticeably better agreement for the remnant BH mass, although slightly less good one for the spin.\\

The comparison for our BNS merger scenarios is more challenging. While Ref.~\cite{Coughlin:2018fis} proposed fitting formulas for the BH remnant mass and spin for BNS mergers, these fits have been calibrated to a smaller parameter space, in particular, to non-spinning NSs. Nevertheless, we list the values for the remnant BH mass and dimensionless spin\footnote{We find the fitting parameters for Eq.~(D7) of Ref.~\cite{Coughlin:2018fis} to be incorrect. Therefore, we have refitted the data and obtained $a =8.58471$, $b=-0.00046$, and $c=-0.51278$. To account for the spins of NSs, we add the individual NS spins to the spin of the remnant BH.} obtained using the fit in Table~\ref{table:remnant_properties_fits}. In comparison, using the ~\mbox{\textsc{NRSur7dq4Remnant}}~\cite{Varma:2018aht,Varma:2019csw,vijay_varma_2018_1435832} surrogate model (with tidal effects neglected) results in a better agreement with our NR results in the spin value, with up to a 9\% difference, and similar agreement for the final remnant BH mass, about 1-2\%.

\subsection{Ejecta and disk}
\label{section:ejecta_and_disk}
The largest amount of ejecta is produced in the BHNS configurations even with the BH having retrograde spin -- the lower spins of the BH are known to reduce the ejecta mass, as confirmed by NR simulations~\cite{Kawaguchi:2015bwa,Foucart:2019bxj,Chen:2024ogz,Martineau:2024zur,Matur:2025avh}.
For the BNS case, the configurations where both NSs are spinning eject more matter than the configuration with high spin only for the primary NS. As is generally the case when tidal ejecta dominate, the configurations with the stiffer EOS produce more ejecta due to lower compactness~\cite{Bauswein:2013yna,Sekiguchi:2016bjd,Palenzuela:2015dqa}. The difference of orders of magnitude in the ejecta mass shows that the outcome properties are highly sensitive to the EOS, spins, and, most of interest here, to the binary type.

We compare the BHNS dynamical ejecta mass with the predictions of the fitting formula in Ref.~\cite{Kruger:2020gig}. For both of our BHNS configurations, the formula yields no dynamical ejecta. The main reason behind this is the high uncertainties, of around $0.01~\mathrm{M}_\odot$, of the fit at low dynamical ejecta masses. Moreover, the fitting was calibrated using NR simulations with $\chi_\mathrm{eff} \in [0, 0.75]$ and $q \in [1/3, 1/7]$, and the parameters of our configurations lie outside of this region.

For the BNS cases, no fitting formula is available for the dynamical ejecta mass for systems where the NSs are rapidly rotating or calibrated using high NS masses, as the ones considered here. To highlight the differences between our NR simulations and the existing models, we apply the formula for non-spinning BNS from Ref.~\cite{Kruger:2020gig}, which produces zero dynamical ejecta mass for all the BNS configurations here. 

We also compare the baryonic mass of the remnant for the BHNS systems using the fitting formula of Ref.~\cite{Foucart:2018rjc}. For BHNS\_DD2, it yields $M_{\mathrm{b}}^{\mathrm{rem}} \simeq 3 \times 10^{-3}$, which is around 10 times less than in our simulations. These results are in good agreement with Ref.~\cite{Martineau:2024zur} for GW230529-like BHNS NR simulations, where the authors also find a similar underprediction of the fitting formula. For BHNS\_MPA1, it yields $M_{\mathrm{b}}^{\mathrm{rem}} = 0$, a value smaller than the one seen in the NR data, mostly because of the high errors for low remnant masses.

\subsection{Kilonova}
We perform Monte Carlo radiative transfer simulations of the ejecta profiles using  \textsc{POSSIS}~\cite{Bulla:2019muo,Bulla:2022mwo} to generate kilonova lightcurves. The code uses heating rates of Ref.~\cite{Rosswog:2022tus}, thermalization efficiencies of Ref.~\cite{Wollaeger:2017ahm}, and wavelength- and time-dependent opacities of Ref.~\cite{Tanaka:2019iqp}, which depend on local properties of the ejecta as density, temperature, and electron fraction. 

The ejecta profiles are obtained from the NR simulations at the latest points of the evolution, which correspond to 12.57~ms and 10.05~ms after merger for BHNS\_DD2 and BHNS\_MPA1, respectively. The BNS configurations did not produce enough ejecta to lead to substantial kilonova emission, so we consider only the BHNS configurations here. We emphasize that we consider only dynamical ejecta due to the omission of modelling of neutrino emission and magnetic fields in our simulations. These processes are responsible for the production of wind ejecta~\cite{Fernandez:2014bra,Kiuchi:2015qua}, and their inclusion would have thus increased the total ejecta mass. Therefore, we anticipate our kilonova lightcurves to be underestimated. We list the values for the dynamical ejecta masses in the second column of Tab.~\ref{table:remnant_properties}. Within \textsc{POSSIS}, the ejecta profiles are rescaled as a function of time, assuming homologous expansion. As the simulations did not include a detailed microphysics of neutrinos, the values for the electron fraction $Y_e$ are inferred using the entropy indicator $\hat{S}$, as outlined in Refs.~\cite{Neuweiler:2022eum,Markin:2023fxx}. We use the threshold $\hat{S}_\mathrm{th} = 50$, and set the electron fraction to $Y_e=0.15$ for $\hat{S} < \hat{S}_\mathrm{th}$, and  $Y_e=0.3$ for $\hat{S} > \hat{S}_\mathrm{th}$. We find that the ejecta in the BHNS cases is almost entirely unshocked, and has mass-weighted electron fraction $\langle Y_e \rangle \sim 0.15$.

As BHNS mergers are known to produce highly axially asymmetric ejecta profiles, it is essential to relax the corresponding symmetry assumption in \textsc{POSSIS}. Thus, instead of placing observers along a fixed azimuthal angle, we distribute them across the entire emission sphere using the \textsc{HEALPix} sphere discretization~\cite{Gorski:2004by} with resolution of $N_{\mathrm{side}}=3$, corresponding to 108 observers.

\begin{figure*}[!htp]
    \centering
    \includegraphics[width=1\linewidth]{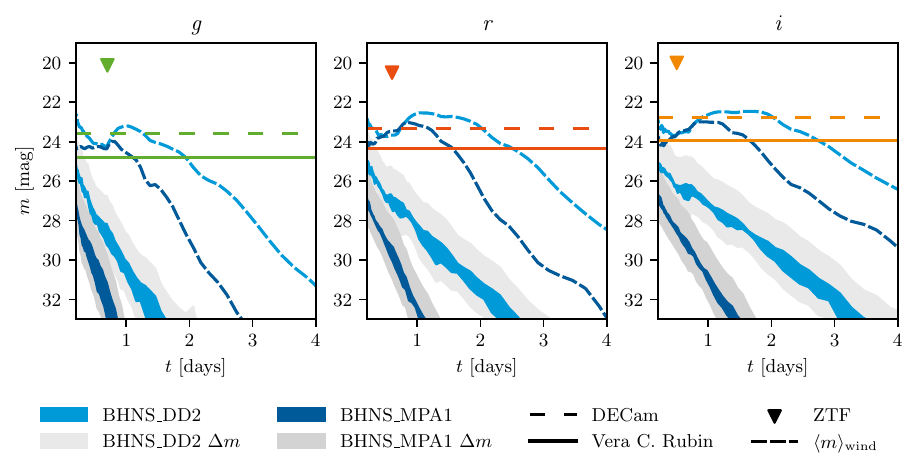}
    \caption{The apparent-magnitude bands containing kilonova lightcurves in \textit{g, r,} and \textit{i} filters for the BHNS configurations averaged over azimuthal angles. The gray bands represent the lightcurve magnitude uncertainties $\Delta m$ induced by the uncertainties in the distance. The colored triangles mark the limiting magnitudes of ZTF follow-up searches of GW230529 in the corresponding filters. The colored horizontal lines show the 5$\sigma$ depth of wide-field searches in the corresponding filters by Vera C. Rubin Observatory and DECam. The average kilonova lightcurves for the models with an additional spherical disk wind component $\langle m \rangle_{\mathrm{wind}}$ are shown as dashed lines.}
    \label{figure:lightcurves_rgi}
\end{figure*}

In Figure~\ref{figure:lightcurves_rgi}, we show the bands of the possible kilonova lightcurves produced by the radiative transfer simulations in SDSS $g$, $r$, and $i$ bands~\cite{Fukugita:1996qt}, along with the allowed lightcurve ranges set by the distance uncertainties. The bands are created by averaging over the azimuthal angles and taking the minimum and maximum magnitudes among polar angles. The kilonova emission originating from the dynamical ejecta peaks at $\sim 25$~mag and rapidly fades over the course of several days, depending on the filter and EOS. The low $Y_e$ of the ejecta also contributes to the low luminosity of the kilonova due to higher opacities of the matter. Compared to BHNS\_MPA1, the BHNS\_DD2 configuration produces a brighter and longer-lasting emission primarily due to the ejecta being 10 times more massive.

In the same Figure~\ref{figure:lightcurves_rgi}, we show the limiting magnitudes of Zwicky Transient Facility (ZTF) of the follow-up searches of transients associated with GW230529~\cite{2023GCN.33900....1K,Ahumada:2024qpr}, and the design \mbox{5-$\sigma$} wide-field depths of DECam~\cite{DES:2018gui} and Vera C. Rubin Observatory~\cite{LSST:2008ijt}. Even with the modern and upcoming wide-field surveys, the kilonova emission would have been too dim to be found. Moreover, the distance to the source plays a significant role -- if the source were located at the distance of GW170817~\cite{LIGOScientific:2017vwq}, i.e., 40~Mpc, it would have been around 3~mag brighter, although detectable by Vera C. Rubin Observatory only during the first day after the merger.

Additionally, we estimate the change in the kilonovae brightness if the wind ejecta were modelled. Assuming that 30\% of the disk mass becomes unbound as wind ejecta~\cite{Nakar:2019fza}, we perform radiative transfer simulations with an additional spherical ejecta component with a constant $Y_e = 0.3$, mass-weighted velocity $\bar{v}=0.05$~\cite{Fernandez:2014bra}, and a power-law matter distribution following Ref.~\cite{Bulla:2020jjr}. We show the average lightcurves from these configurations in Figure~\ref{figure:lightcurves_rgi} (dashed lines). Due to the additional wind ejecta, the kilonova emission becomes noticeably brighter and is roughly visible for $\sim 2$ days after the merger. This brightness increase highlights the importance of detailed modelling of post-merger disk wind ejecta driven by the neutrino emission, magnetic fields, and viscosity, e.g.,~\cite{Fernandez:2014bra,Kiuchi:2015qua,Fernandez:2018kax,Nedora:2020hxc}.

In Ref.~\cite{Kunnumkai:2024qmw}, they find GW230529 to produce a brighter kilonova than in our case, peaking at $g \lesssim 23.5$ and $i < 23$. Dimmer kilonovae in our simulations are the result of the absence of wind ejecta in our simulations. In Ref.~\cite{Kunnumkai:2024qmw}, the wind ejecta is included in the model by construction.

The population-synthesis study of the origins of GW230529 in Ref.~\cite{Chandra:2024ila} finds the GW230529 kilonova to have slightly higher but similar brightness as in our study. There, the wind ejecta mass was estimated using the fitting formula of Ref.~\cite{Raaijmakers:2021slr}, which, in turn, relies on the baryonic remnant mass fit from Ref.~\cite{Kruger:2020gig}. As shown previously in Sec.~\ref{section:ejecta_and_disk}, the latter fit underpredicts the amount of the baryonic remnant mass by an order of magnitude for the BHNS systems with mass ratios similar to that of GW230529. However, they assume the BH to have zero spin, which leads to an overestimation of the baryonic remnant mass by a factor of 10 or larger for their highest-likelihood BHNS configuration compared to our case, where the BH has spin $\chi_\mathrm{BH}=-0.11$. These factors appear to roughly cancel each other out, explaining similar estimates of the kilonova brightness.
In Ref.~\cite{Pillas:2025pfc}, the same fitting formulae were used to estimate the ejecta mass, and, thus, are subject to the limitations described above. Consistent with our results, they conclude that the expected kilonova emission was likely too dim to be located by the current wide-field surveys. However, the Vera C. Rubin Observatory was not operating at the time of the GW230529 event, although it might have been instrumental for its detectability.

\section{Conclusion}
\label{section:conclusion}
In this paper, we have studied two possible scenarios for GW230529, namely, a lower-mass-gap BHNS merger or a heavy, rapidly spinning BNS merger. We use NR simulations to produce numerical-relativity waveforms that accurately describe the late inspiral and the merger phase. The numerical waveforms show overall good agreement with the BNS waveform model of choice, \verb|IMRPhenomXAS_NRTidalv3|. This demonstrates the robustness of the waveform model even for our BHNS cases and for BNS configurations that have large component masses well outside of the calibration region.

Combining these numerical-relativity and waveform model data, we produce hybrid waveforms (hybrids) covering the detection frequency range of Advanced LIGO~\cite{LIGOScientific:2014pky}. Using the detector sensitivity at the time of the event, we performed PE on the injected hybrids.

Most importantly, we find that the posterior distributions for the effective spin show a different grouping depending on the type of injection, i.e., a BNS or BHNS system. When we compare the posteriors of the injected waveforms and the ones obtained by reanalysing GW230529, there is an overall larger agreement between the BHNS injections and GW230529. This agreement can be quantified by computing the Jensen-Shannon divergence for posteriors in mass ratio and effective spin. Under different waveform models (with or without tides, with or without spin precession), and spin priors, the posteriors closest to GW230529 are the ones with aligned-spin models with tides. 

Based on the posterior similarities, we conclude that GW230529 was more likely a BHNS merger event rather than a BNS one. However, the BNS merger origin cannot be ruled out.

Analyzing the remnant BH properties, we find that the BHNS cases leave behind more massive and slower-spinning remnant BHs compared to the BNS systems. For comparison with our results, we employ various fitting formulae for the remnant properties, and find that our NR results are in excellent agreement with the fitting formula for the remnant BH mass and spin in Ref.~\cite{Gonzalez:2022prs}. In contrast to this, the analytical predictions for the dynamical ejecta mass predict no ejecta produced in all configurations, while we obtain a non-negligible ejecta mass for the BHNS configurations. We find that the fitting formula for the baryonic mass of BHNS remnants overpredicts it by a factor of ten in the case of stiffer EOS, and underpredicts it for the softer EOS. The same result was found by~\cite{Martineau:2024zur}. spins and NSs with higher mass for higher-quality fits in this region of the parameter space.

For the BHNS systems, as they produce the highest amount of ejecta, we perform radiative transfer simulations using the dynamical ejecta profiles from our NR data. We find the kilonova emission peaking at magnitudes just outside the reach of the current and upcoming wide-field surveys. Low masses of the dynamical ejecta, together with the large distance of the source, play the decisive role for such an event to be undetectable. However, the inclusion of a simple analytical prescription for wind ejecta is capable of significantly boosting the kilonova emission, making it detectable by the Vera C. Rubin Observatory during the first two days after merger.

\section{Data availability}
We publish the following data products: the initial data~\cite{markin_2025_16843186}, the numerical waveforms~\cite{markin_2025_16875324}, and the posterior samples~\cite{markin_2025_16874963}. Further data products of this study can be obtained upon a reasonable request to the corresponding author.

\section{Acknowledgements}
We are grateful to Natalie Williams, Colm Talbot, Geraint Pratten, and Frank Ohme for providing useful insights into parameter estimation; to Samuel Tootle for investigating diverging initial data; to Adrian Abac for sharing valuable insights into the waveform models; to Marta Colleoni for reviewing the manuscript within LVK; and to the anonymous referee for providing valuable feedback, which improved our paper.
I.M. and T.D. gratefully acknowledge support from the Deutsche Forschungsgemeinschaft (DFG) under the project 504148597 (DI 2553/7).  T.~D. and A.~P. acknowledge funding from the European Union (ERC, SMArt, 101076369). 
Views and opinions expressed are those of the authors only and do not necessarily reflect those of the European Union or the European Research Council. Neither the European Union nor the granting authority can be held responsible for them. M.B. acknowledges the Department of Physics and Earth Science of the University of Ferrara for the financial support through the FIRD 2024 grant. The authors gratefully acknowledge the computing time granted by the Resource Allocation Board and provided on the supercomputer Emmy/Grete at NHR-Nord@Göttingen as part of the NHR infrastructure. The calculations for this research were conducted with computing resources under the project bbp00049.
The PE was done on the DFG-funded research cluster Jarvis at the University of Potsdam (INST 336/173-1; project number: 502227537).
This research has made use of data or software obtained from the Gravitational Wave Open Science Center (\url{gwosc.org}), a service of the LIGO Scientific Collaboration, the Virgo Collaboration, and KAGRA. This material is based upon work supported by NSF's LIGO Laboratory which is a major facility fully funded by the National Science Foundation, as well as the Science and Technology Facilities Council (STFC) of the United Kingdom, the Max-Planck-Society (MPS), and the State of Niedersachsen/Germany for support of the construction of Advanced LIGO and construction and operation of the GEO600 detector. Additional support for Advanced LIGO was provided by the Australian Research Council. Virgo is funded, through the European Gravitational Observatory (EGO), by the French Centre National de Recherche Scientifique (CNRS), the Italian Istituto Nazionale di Fisica Nucleare (INFN) and the Dutch Nikhef, with contributions by institutions from Belgium, Germany, Greece, Hungary, Ireland, Japan, Monaco, Poland, Portugal, Spain. KAGRA is supported by Ministry of Education, Culture, Sports, Science and Technology (MEXT), Japan Society for the Promotion of Science (JSPS) in Japan; National Research Foundation (NRF) and Ministry of Science and ICT (MSIT) in Korea; Academia Sinica (AS) and National Science and Technology Council (NSTC) in Taiwan.

\appendix

\section{Convergence of numerical waveforms}
\label{section:convergence}
To assess the accuracy of the numerical simulations, we perform convergence tests for the extracted gravitational waveforms.

\begin{figure}[htp]
    \centering
    \includegraphics[width=1\linewidth]{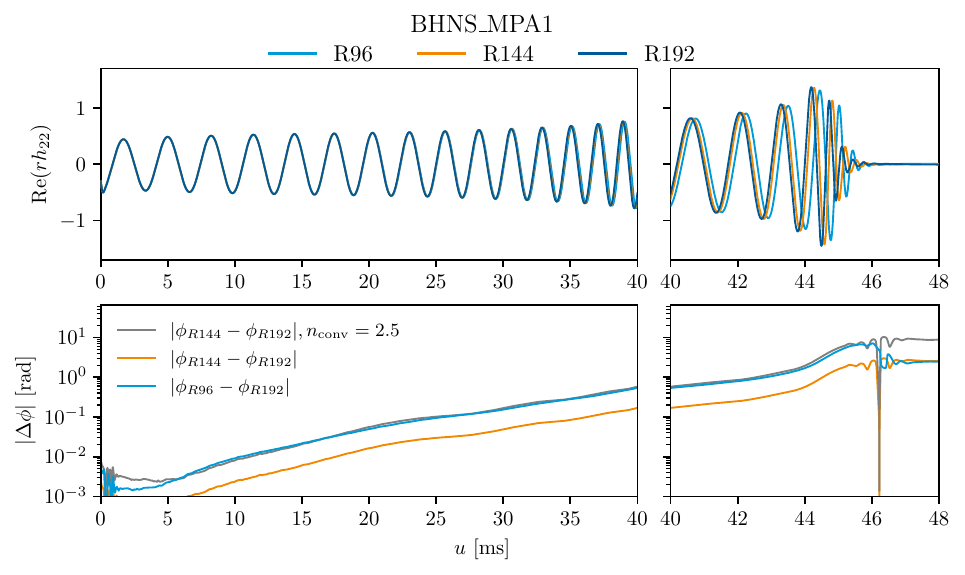}
    \includegraphics[width=1\linewidth]{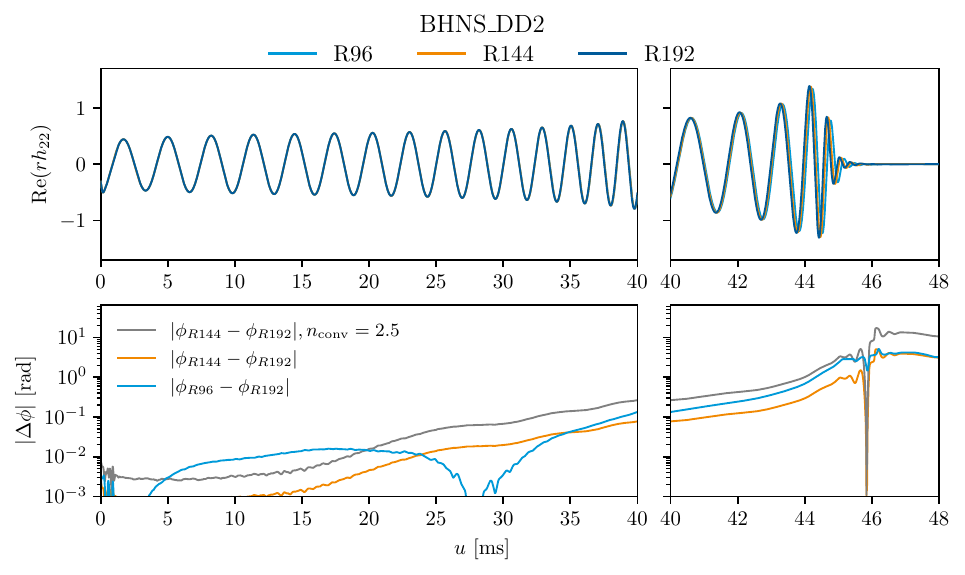}
    \includegraphics[width=1\linewidth]{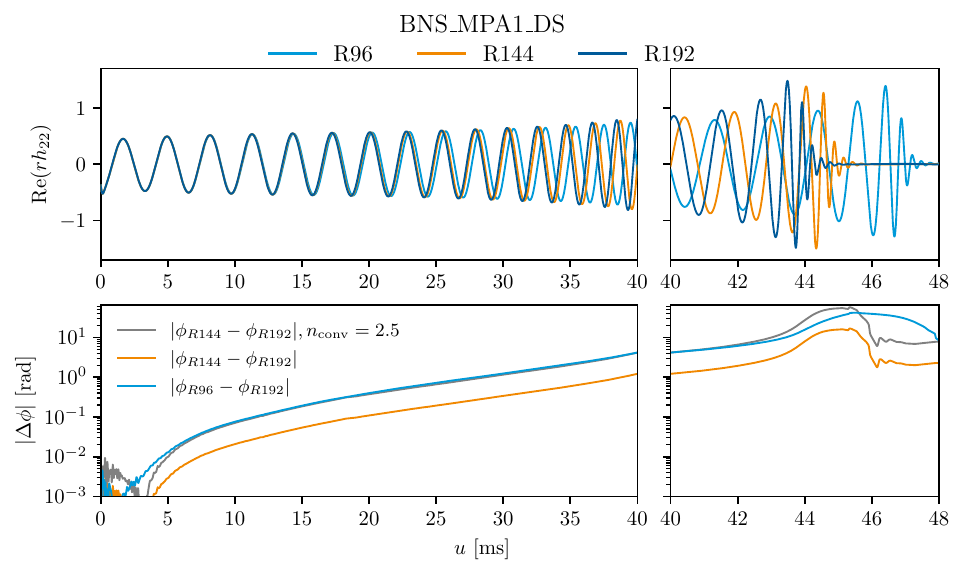}
    \caption{Comparison of the numerical waveforms at different resolutions for a selection of configurations: BHNS\_MPA1, BHNS\_DD2, and BNS\_MPA1\_DS. \textit{Upper panels}: real part of the waveform evaluated on the largest extraction radius for each resolution. \textit{Bottom panels}: The phase differences between the lowest resolutions and the highest one, with the gray line representing rescaled dephasing assuming convergence order $n_{\mathrm{conv}}$.}
    \label{figure:nr_waveforms_convergence}
\end{figure}

In Figure~\ref{figure:nr_waveforms_convergence}, we show the waveforms at all three resolutions: R96, R144, R192, and the phase differences between the two lowest resolutions and the highest one\footnote{For some runs, a small amount of the waveform data was lost due to technical reasons. We fill the short gaps in the waveform data using cubic spline interpolation.}.

Similar to previous studies~\cite{Foucart:2013psa,Chaurasia:2021zgt}, in the case of BHNS mergers, lower resolutions delay the merger relative to the higher ones. The BHNS\_MPA1 waveform shows an overall convergence order of $\sim 2.5$. For BHNS\_DD2, there appear to be multiple competing error contributions, with alternating influences during different stages of the inspiral, which result in the zero crossing of the lowest-highest resolution dephasing around 27~ms. Thus, the apparently good agreement of the merger times might be coincidental and does not imply a significantly higher convergence order. As this effect depends on the EOS, it is crucial to verify convergence across various physical setups to select an appropriate resolution.

The BNS configurations, however, exhibit the same reverse trend compared to non-spinning BNSs~\cite{Radice:2013hxh,Dietrich:2019kaq}, with high aligned spins~\cite{Kuan:2025bzu}, or with moderate antialigned spins ($\chi = -0.1$)~\cite{Schianchi:2024vvi}. Here, the merger happens at the latest for the lowest resolution, with waveforms converging to the ones with earlier merger times. A similar situation can be seen in Figure~4d of~\cite{Dudi:2021wcf} for a system with $\chi=-0.28$, where the lowest resolution has the latest merger time. However, for the other resolutions in the study, the waveforms switch to the opposite trend. Further convergence studies based on a larger set of simulations at different resolutions are required to make conclusive statements. We note that the dephasing between the lowest resolution and the highest one is significantly larger in BNS configurations. With such large dephasings, it is crucial to perform simulations at higher resolution to accurately model the BNS systems. Overall, the BNS waveforms show a convergence of $\sim2.5$ order throughout almost the entire inspiral. 

\section{Impact of different sampler settings}
\label{section:sampler_settings}
To verify how the employed sampler settings affect the results of PE, we run PE with the following changes in sampler settings, which are summarized in Table~\ref{table:sampler_settings}; technical details about the different settings and their options can be found in Ref.~\cite{Speagle:2019ivv} and references therein.

\textit{Standard settings}: We set 2000 live points, with the standard stopping criterion $d\log{\mathcal{Z}}=0.1$. The bounding strategy is set to the unit N-cube (no bounding), i.e., the samples are drawn directly from the prior (\texttt{bound=live} setting). 

\textit{Higher number of live points}: Increasing the number of live points can allow for higher accuracy of the evidence and posterior estimates by increasing the coverage of the parameter space. This might be especially helpful if multiple modes are present. In this setting, we set the number of live points to 4000.

\textit{Multi-ellipsoid bound}: Multi-ellipsoid bound provides higher sampling efficiency and better captures multi-modal distributions if they are present. It can also have higher resolution in the tails of the distribution. In contrast with no bounding in the standard settings, here, we change the bound to a multi-ellipsoid bound (\texttt{bound=live-multi}). 

\textit{Stricter stopping criterion}: By default, \textsc{Bilby} stops the nested sampling algorithm when the estimated remaining contribution to the log-evidence $d\log{\mathcal{Z}}$ reaches $0.1$. Often, this is a reasonable choice, as it means that the sampler has already explored most of the parameter space and there is only a small estimated fraction of evidence that is unaccounted for. However, decreasing  $d\log{\mathcal{Z}}$ makes the sampling run for longer, potentially providing more accurate posterior estimates by allowing the sampler to explore the high-likelihood regions more. In this work, we decrease the stopping criterion to $d\log{\mathcal{Z}}=10^{-3}$.

\begin{table}[!htp]
    \centering
    \begin{tabular}{l|ccc}
        Setting & $n_{\mathrm{live}}$ & Bound & $d \log{\mathcal{Z}}$ \\
        \hline
        \textit{Standard settings} & 2000 & N-cube & 0.1 \\
        \textit{Higher number of live points} & 4000 & N-cube & 0.1 \\
        \textit{Multi-ellipsoid bound} & 2000 & Multi-ellipsoid & 0.1 \\
        \textit{Stricter stopping criterion} & 2000 & N-cube & 0.001 \\

    \end{tabular}
    \caption{Summary of the tested sampler settings. The columns list the setting name, number of live points $n_{\mathrm{live}}$, bounding strategy, and stopping criterion  $d \log{\mathcal{Z}}$, respectively.}
    \label{table:sampler_settings}
\end{table}

\begin{figure*}[htp]
    \centering
    \includegraphics[width=0.48\linewidth]{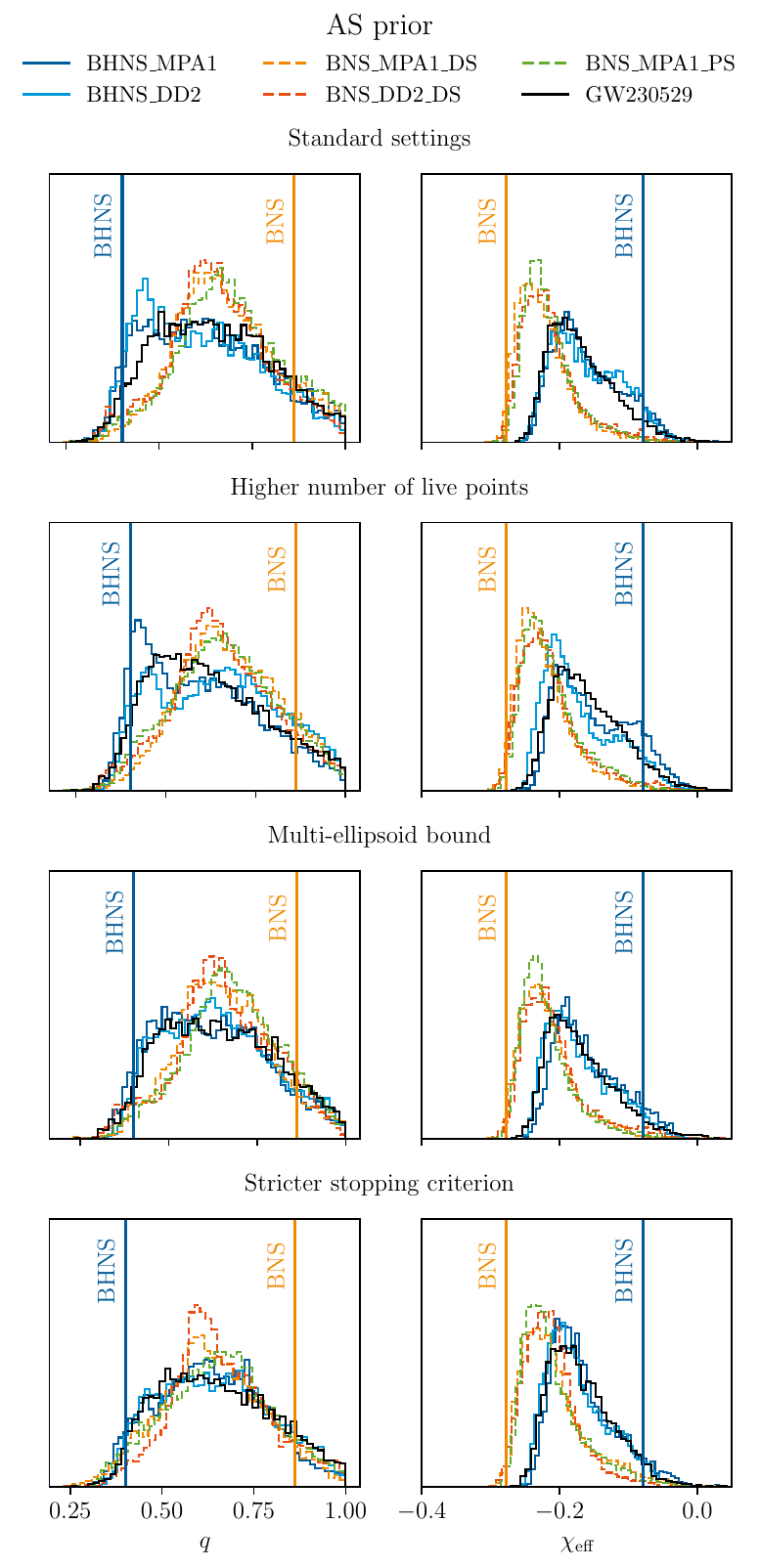}
    \includegraphics[width=0.48\linewidth]{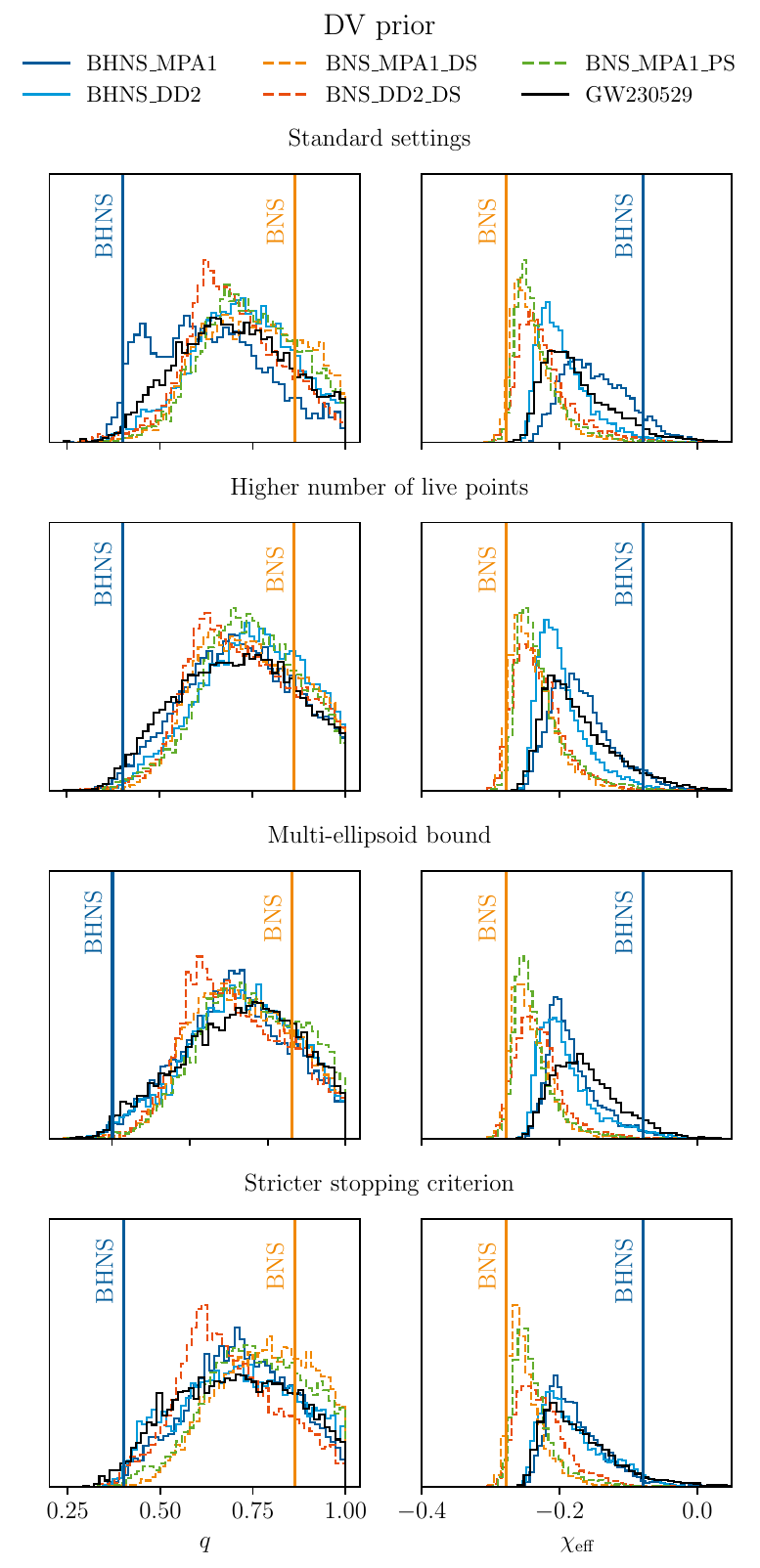}
    \caption{Posterior samples for mass ratio and effective spin for all injections and the GW230529 event data under different sampler settings (\textit{panel row titles}) under the AS prior (\textit{left columns}) and the DV prior (\textit{right columns}), both recovered using the \texttt{IMRPhenomXAS\_NRTidalv3} waveform model.}
    \label{figure:sampler_settings}
\end{figure*}

In Figure~\ref{figure:sampler_settings}, we show the results for the reanalysis of both injections and GW230529 data using the different settings described above. The higher number of live points has little impact on the posteriors in the case of the AS prior, while making the mass ratio posteriors more similar to each other and the effective spin posteriors more peaked for the DV prior. The multi-ellipsoid bounding also has a minor impact on the posteriors, only making the BHNS posteriors agree with each other and the GW230529 one under the AS prior.

The stricter stopping criterion yields the most significant changes to the posteriors. For both AS and DV priors, the mass ratio posteriors start to agree well with each other regardless of the type of injection and the data used; the posteriors for the effective spin exhibit a high level of agreement among the source groups, BHNS and BNS. The posterior distributions of the BHNS and GW230529 cases are largely overlapping. The stricter stopping criterion thus makes the distinction between the two binary types clearer, regardless of the prior employed.

We conclude that among the adjusted sampler settings, the highest accuracy of the posteriors is achieved using a stricter stopping criterion. For that reason, we chose this setting in the main analysis.

\section{Carbon footprint}
\label{section:carbon_footprint}
Performing high-resolution NR simulations requires a large amount of energy. Most of the time, computing facilities are attached to national grids, which can acquire energy from a mix of different sources, including fossil fuels. This in turn induces a tangible amount of greenhouse gas (GHG) emissions, which are driving climate change~\cite{Oreskes_2004,Doran_2009,Cook_2013,Cook_2016,Lynas_2021,Myers_2021}.

Here, we quantify the GHG emissions footprint of this study. During the runtime, we collect the utilization CPU metrics using \textsc{calcicum}~\cite{calcium}, which automatically obtains the Thermal Design Power (TDP) of the CPU model, and estimates the energy consumed by the computations.

We list the energy consumption and the corresponding greenhouse gas emissions in Table~\ref{table:carbon_footprint}. The energy estimates only account for the CPU usage and not that of memory and network components. To account for cooling and other operational overheads, we adopt Power Usage Effectiveness (PUE)~\cite{YUVENTI201390} to be $1.03$ for Emmy~\cite{NHRGoettingenPUE} and $1.1$ for Jarvis~\cite{hpc_energy} and rescale the CPU energy and emission data. Since Emmy supercomputer switched to electricity sourced completely from renewable sources in 2022~\cite{NHRGoettingenPUE}, we quote the avoided emissions calculated using the average carbon intensity of the German electricity grid.

\begin{table}[!htbp]
    \centering
    \begin{tabular}{l|c|c|c}
        & Energy & Emissions, &  Avoided emissions,\\
        & MWh & t\ch{CO2}e & t\ch{CO2}e\\
        \hline
Evolution R192 & 14.92 & -- & 5.68 \\
Evolution R144 & 5.00 & -- &  1.93 \\
Evolution R96 & 2.07 & -- & 0.79 \\
Eccentricity reduction & 1.03 & -- & 0.39 \\
Parameter estimation & 0.182 & 0.069 & -- \\
Radiative transfer & 0.004 & 0.002 & -- \\ 
\hline
Total & 23.26 & 0.071 & 8.79 \\
    \end{tabular}
    \caption{Energy consumption for each step of this study and the corresponding estimated produced and avoided \ch{CO2}e emissions.}
    \label{table:carbon_footprint}
\end{table}

\bibliography{main}

\end{document}